\documentclass[twocolumn,linenumbers]{aastex631}

\begin{document}
\nolinenumbers
\title{
The peculiar disk evolution of 4U 1630--472 observed by Insight-HXMT during the 2022 and 2023 outbursts}

\author[0000-0002-5554-1088]{Jing-Qiang Peng\textsuperscript{*}}
\email{pengjq@ihep.ac.cn}
\affiliation{Key Laboratory of Particle Astrophysics, Institute of High Energy Physics, Chinese Academy of Sciences, 100049, Beijing, China}
\affiliation{University of Chinese Academy of Sciences, Chinese Academy of Sciences, 100049, Beijing, China}
\author{Shu Zhang\textsuperscript{*}}
\email{szhang@ihep.ac.cn}
\affiliation{Key Laboratory of Particle Astrophysics, Institute of High Energy Physics, Chinese Academy of Sciences, 100049, Beijing, China}

\author[0000-0001-5160-3344]{Qing-Cang Shui\textsuperscript{*}}
\email{shuiqc@ihep.ac.cn}
\affiliation{Key Laboratory of Particle Astrophysics, Institute of High Energy Physics, Chinese Academy of Sciences, 100049, Beijing, China}
\affiliation{University of Chinese Academy of Sciences, Chinese Academy of Sciences, 100049, Beijing, China}

\author[0000-0001-8768-3294]{Yu-Peng Chen}
\affiliation{Key Laboratory of Particle Astrophysics, Institute of High Energy Physics, Chinese Academy of Sciences, 100049, Beijing, China}

\author[0000-0001-5586-1017]{Shuang-Nan Zhang}
\affiliation{Key Laboratory of Particle Astrophysics, Institute of High Energy Physics, Chinese Academy of Sciences, 100049, Beijing, China}
\affiliation{University of Chinese Academy of Sciences, Chinese Academy of Sciences, 100049, Beijing, China}

\author[0000-0003-3188-9079]{Ling-Da Kong}
\affiliation{Institute f{\"u}r Astronomie und Astrophysik, Kepler Center for Astro and Particle Physics, Eberhard Karls, Universit{\"a}t, Sand 1, D-72076 T{\"u}bingen, Germany}
\author{A. Santangelo}
\affiliation{Institute f{\"u}r Astronomie und Astrophysik, Kepler Center for Astro and Particle Physics, Eberhard Karls, Universit{\"a}t, Sand 1, D-72076 T{\"u}bingen, Germany}
\author{Zhuo-Li Yu}
\affiliation{Key Laboratory of Particle Astrophysics, Institute of High Energy Physics, Chinese Academy of Sciences, 100049, Beijing, China}
\author[0000-0001-9599-7285]{Long Ji}
\affiliation{School of Physics and Astronomy, Sun Yat-Sen University, Zhuhai, 519082, China}

\author[0000-0002-6454-9540]{Peng-Ju Wang}
\affiliation{Institute f{\"u}r Astronomie und Astrophysik, Kepler Center for Astro and Particle Physics, Eberhard Karls, Universit{\"a}t, Sand 1, D-72076 T{\"u}bingen, Germany}

\author[0000-0003-4856-2275]{Zhi Chang}
\affiliation{Key Laboratory of Particle Astrophysics, Institute of High Energy Physics, Chinese Academy of Sciences, 100049, Beijing, China}
\author{Jian Li}
\affiliation{CAS Key Laboratory for Research in Galaxies and Cosmology, Department of Astronomy, University of Science and Technology of China, Hefei 230026, China}
\affiliation{School of Astronomy and Space Science, University of Science and Technology of China, Hefei 230026, China}
\author[0000-0003-2310-8105]{Zhao-sheng Li}
\affiliation{ Key Laboratory of Stars and Interstellar Medium, Xiangtan University, Xiangtan 411105, Hunan, China}



\begin{abstract}
\nolinenumbers

We study the spectral properties of the black hole X-ray transient binary 4U 1630--472 during the 2022 and 2023 outbursts with Insight-HXMT observations.  We find that the outbursts are in peculiar soft states.
The effect of the hardening factor on the disk temperature is taken into account by kerrbb, and the flux and temperature of the disk are found to follow $F \propto T_{\rm eff}^{3.92\pm 0.13}$ and $F \propto T_{\rm eff}^{4.91\pm 1.00}$, for the two outbursts respectively. The flux-temperature relation is roughly consistent with holding a standard disk,
By fitting with the p-free model, the p-value is found to have anti-correlation with disk temperature. Combined a joint diagnostic in a diagram of the relation between the non-thermal fraction and luminosity, by enclosing as well the previous outbursts, reveals a possible pattern for the disk evolution toward a slim one, and such an evolution may depend on the fraction of the non-thermal emission in the high soft state.

\end{abstract}



\section{Introduction} \label{sec:intro}

Black hole X-ray binaries (BHXRBs)  can be classified as persistent or transient sources. For transient sources, both the accretion rate and the disk temperature are relatively low during long periods in the quiescent state \citep{2009Deegan,2016Tetarenko}.
When the temperature of the disk increases, neutral hydrogen within the outer part of the disk becomes ionized, triggering thermal and viscous instabilities. Consequently, the black hole X-ray binary undergoes an outburst, accompanied by the increase of X-ray emission \citep{1995Cannizzo,2001Lasota,2011Belloni}.

The outbursts of black hole X-ray binaries usually go through different spectral states. Different spectral states will have different spectral and timing characteristics.
For the Low/Hard states (LHS), the emission is dominated by non-thermal emission. Occasionally, it will be accompanied by low-frequency quasi-periodic oscillations (LFQPOs), mainly C-type. The High/Soft state (HSS) emission is dominated by thermal radiation from the disk, characterized additionally by weak noise components in the power density spectrum (PDS). As the accretion rate rises, the black hole X-ray binary may enter the very high state (VHS) (also called the steep power-law (SPL) state), during which the spectral index $\Gamma$ is greater than 2.4 and sometimes QPOs of 0.1--30 Hz appear in the PDS \citep{1975Thorne,1996Tanaka,2005Belloni}.

Black hole X-ray binary outbursts are highly driven by the evolution of the accretion disk. 
Accretion models in black hole X-ray binaries evolved into the Shakura-Sunyaev Disk (SSD), Shapiro-Lightman-Eardley  (SLE) disk, slim disk, and advection-dominated accretion flow (ADAF).
SSDs are geometrically thin, optically thick, and referred to as the standard disk.  The structure and radiation of the stabilized accretion disk can be determined by three fundamental parameters: the coefficient of viscosity $\alpha$, the mass of the central star $M$, and the accretion rate $\dot{M}$. For the BHXRBs, the accretion disk is likely truncated at the inner part at a distance far from black hole and is usually considered as a standard one in the soft state \citep{1973Shakura}.
SLE disks are governed by gas pressure and characterized by different temperatures of the ion and electron populations.  SLE disks can generate intense X-ray and gamma-ray radiation but are thermally unstable \citep{1976S,1978P}. 
ADAFs are geometrically thick and optically thin and dominated by gas pressure. They can maintain viscosity and thermal stability \citep{1995N,1995A}, and hence are commonly utilized in modeling the non-thermal emission of the outbursts \citep{1997Esin,2015K}.
Although ADAFs have low radiation efficiency, the high-temperature electrons within them can scatter soft photons from SSDs through the inverse Compton procedure, producing a significant amount of non-thermal emission.
Slim disks, also known as optically thick ADAFs, were introduced to be optically and geometrically thick by \cite{1988A}, with radiation efficiency lower than the standard disk and the disk emission following $L_{\rm disk}\propto T_{\rm in}^{2}$.

4U 1630--472 is a transient low-mass black hole X-ray binary system that was first discovered by the Uhuru satellite in 1969 \citep{1972G}. This system is well known for its X-ray outburst repeating period of 600--650 days \citep{1976J,1997MNRAS.291...81K,2016ApJS..222...15T}.
A high hydrogen absorption column density was observed as $N_{\rm H}= (4-12) \times 10^{22} \rm cm ^{-2}$ \citep{1998T}.
One of the peculiar behaviors for 4U 1630--472 is that the low hard state was hardly caught in the initial outburst phase \citep{2005A,2014T}, similar to outbursts observed in other sources such as MAXI J0637--430 and SLX 1746--331   \citep{2022Ma,2023P,2024P}.
4U 1630--472 was observed to have a black hole mass of about 10 $M_{\odot}$ \citep{2014S}, an inclination of 65° \citep{1998Kuu}, and a distance between 4.7 and 11.5 kpc \citep{2018K}.  For the black hole spin, \cite{2022L} reported $a$ = 0.817 $\pm $ 0.014 using Insight-HXMT data, and \cite{2014K} obtained $0.985_{-0.014}^{+0.005}$  by
fitting the spectrum of NuSTAR.
\cite{2005A}  classified the disk of 4U 1630--472 into three states with a series of outbursts observed in 1996-2006 by RXTE.   IXPE observation of 4U 1630--472 in HSS of the 2022 outburst revealed very high polarization (from $\sim$ 6\% at 2 keV to $\sim$ 10\% at 8 keV), which is not consistent with the standard thin accretion disk model \citep{2024R}. \cite{2024R} suggested that matter is accreted onto the black hole through a thin disk covered with a partially ionized atmosphere that flows away at mildly relativistic speeds.

In this paper, we perform a detailed spectral analysis by taking  Insight-HXMT observations. We report the evolution of the disk during the 2022 and 2023 outbursts. Section \ref{obser} describes the observations and data reduction. The detailed results are presented in Section \ref{result}. The results are then discussed and concluded in Section \ref{dis}.

\section{Observations and Data reduction}
\label{obser}

\subsection{Insight-HXMT}
Insight-HXMT is the first Chinese X-ray astronomy satellite, successfully launched on  2017, June 15 \citep{2014Zhang, 2018Zhang, 2020Zhang}.  It carries three scientific payloads: the low energy X-ray telescope (LE, SCD detector, 1--15 keV, 384 $\rm cm^{2}$, \citealt{2020Chen}), the medium energy X-ray telescope (ME, Si-PIN detector, 5--35 keV, 952 $\rm cm^{2}$, \citealt{2020Cao} ), and the high energy X-ray telescope (HE, phoswich NaI (CsI), 20--250 keV, 5100 $\rm cm^{2}$, \citealt{2020Liu}).

Insight-HXMT observed 4U 1630--472 from August 6, 2022 (MJD 59797) to May 3, 2023 (MJD 60070).
We extract the data from LE, ME and HE using the Insight-HXMT Data Analysis software {\tt{HXMTDAS v2.06}}. 
The data are filtered with the criteria recommended by the Insight-HXMT Data Reduction Guide {\tt v2.06} \footnote{\url{http://hxmtweb.ihep.ac.cn/SoftDoc/648.jhtml}}.  Due to the large detector noise peaks in the low-energy region of the LE, 
the energy bands considered for spectral analysis and light curves are LE 2--8 keV, ME 8--28 keV and HE 28--100 keV. One percent systematic error is added to data \citep{2020Liao}, and errors are estimated via  Markov Chain Monte-Carlo (MCMC) chains with a length of 20000.

\section{Results}
\label{result}

\subsection{Light curve and Hardness-intensity diagram}
\label{light}

\begin{figure*}
\centering

\begin{minipage}[t]{0.45\linewidth}
\centering
\includegraphics[angle=0,scale=0.4]{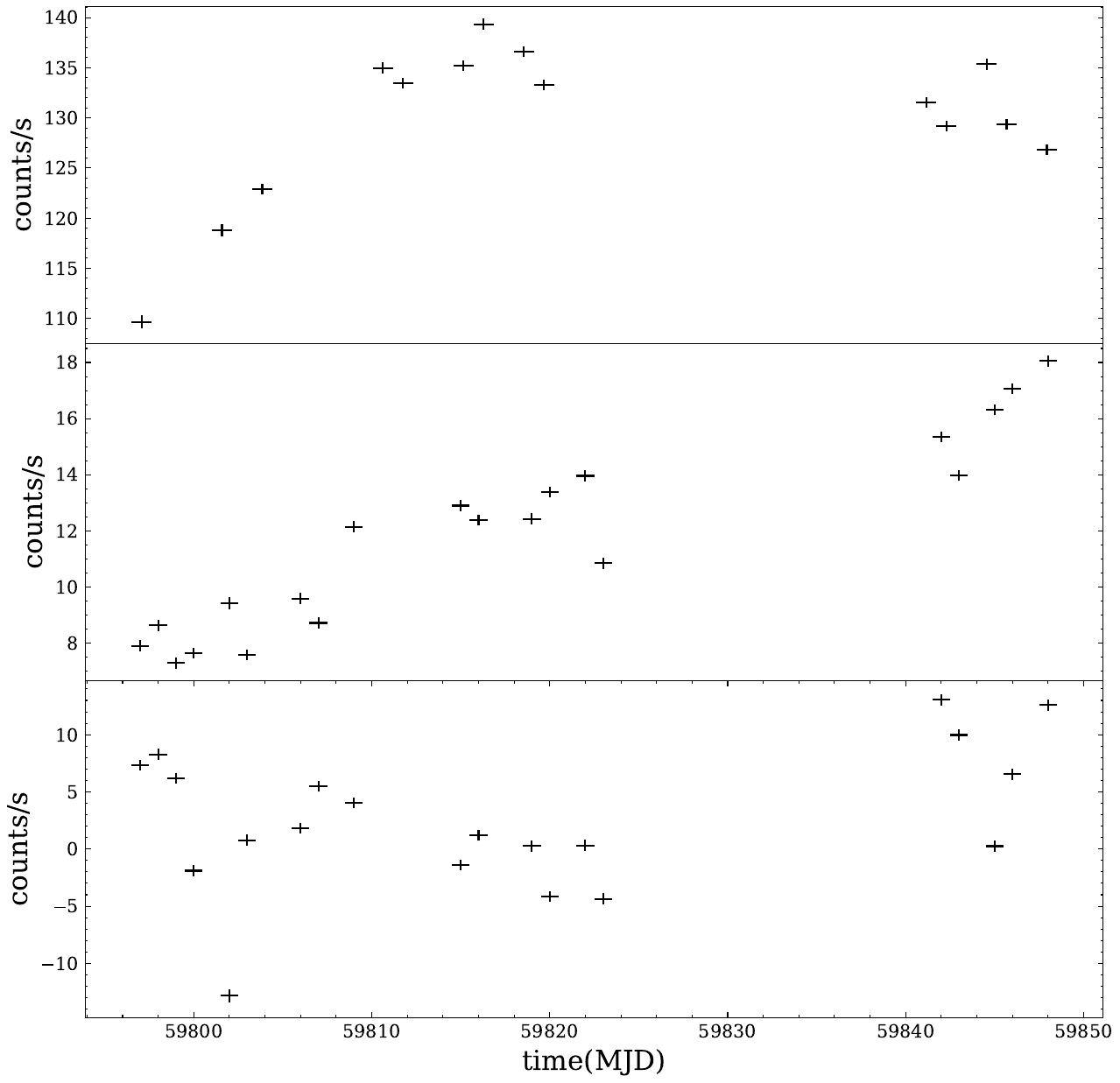}
\end{minipage}%
\hfill
\begin{minipage}[t]{0.45\linewidth}
\includegraphics[angle=0,scale=0.4]{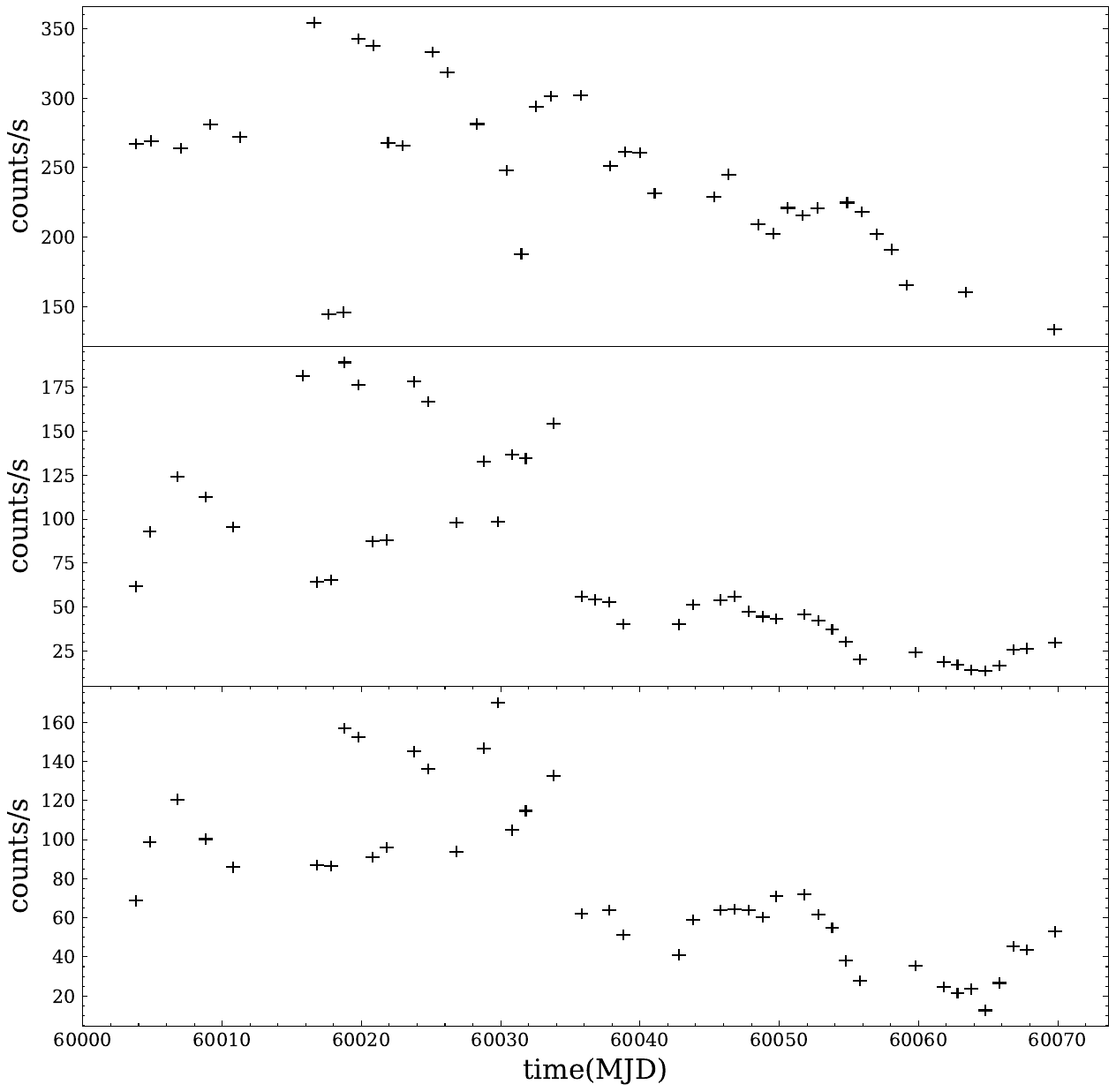}
\end{minipage}
	\caption{ The light curves of 4U 1630--472 observed by Insight-HXMT during 2022 and 2023 outbursts. Top panel: the light curve of Insight-HXMT LE in 2--8 keV.
Middle panel: the light curve of Insight-HXMT ME in 10--28 keV.
Bottom panel: the light curve of Insight-HXMT HE in 28--100 keV. }
	\label{lcurve}
\end{figure*}

\begin{figure}
	\centering
	\includegraphics[angle=0,scale=0.4]{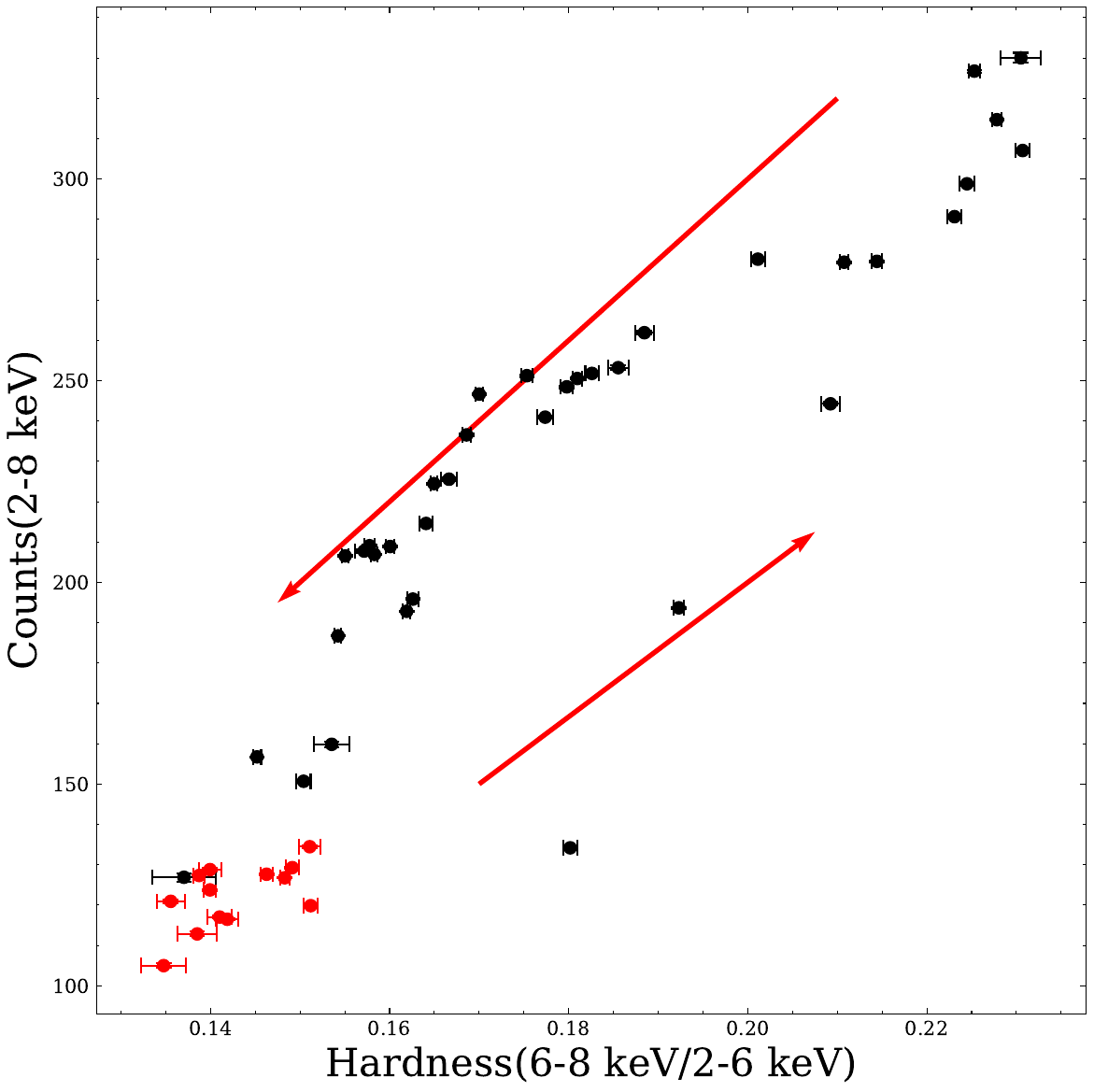}
	\caption{ The hardness-intensity diagram of 4U 1630--472, where the hardness is defined as the ratio of 6--8 keV to 2--6 keV count rate. The black and red dots represent the 2022 and 2023 outbursts of  4U 1630--472 observed by Insight-HXMT. }
	\label{hid}
\end{figure}

Figure \ref{lcurve} shows the light curve of Insight-HXMT observations of 4U 1630--472, where two outbursts stand out in 2022 and 2023, respectively.

We extract the 2--6 keV, 6--8 keV, and 2--8 keV light curves of Insight-HXMT to construct the Hardness-Intensity Diagram (HID) of 4U 1630--472. The hardness is defined as the count rate ratio of 6--8 keV to 2--6 keV, while the intensity takes the count rate of 2--8 keV.  
As shown in Figure \ref{hid}, the HID from Insight-HXMT observations suggests that the source evolved from 2022 to 2023 with a harder spectrum.

\subsection{The spectral analysis}
\label{parameters}

\subsubsection{ diskbb}

\begin{figure*}
\centering

\begin{minipage}[t]{0.45\linewidth}
\centering
\includegraphics[angle=0,scale=0.4]{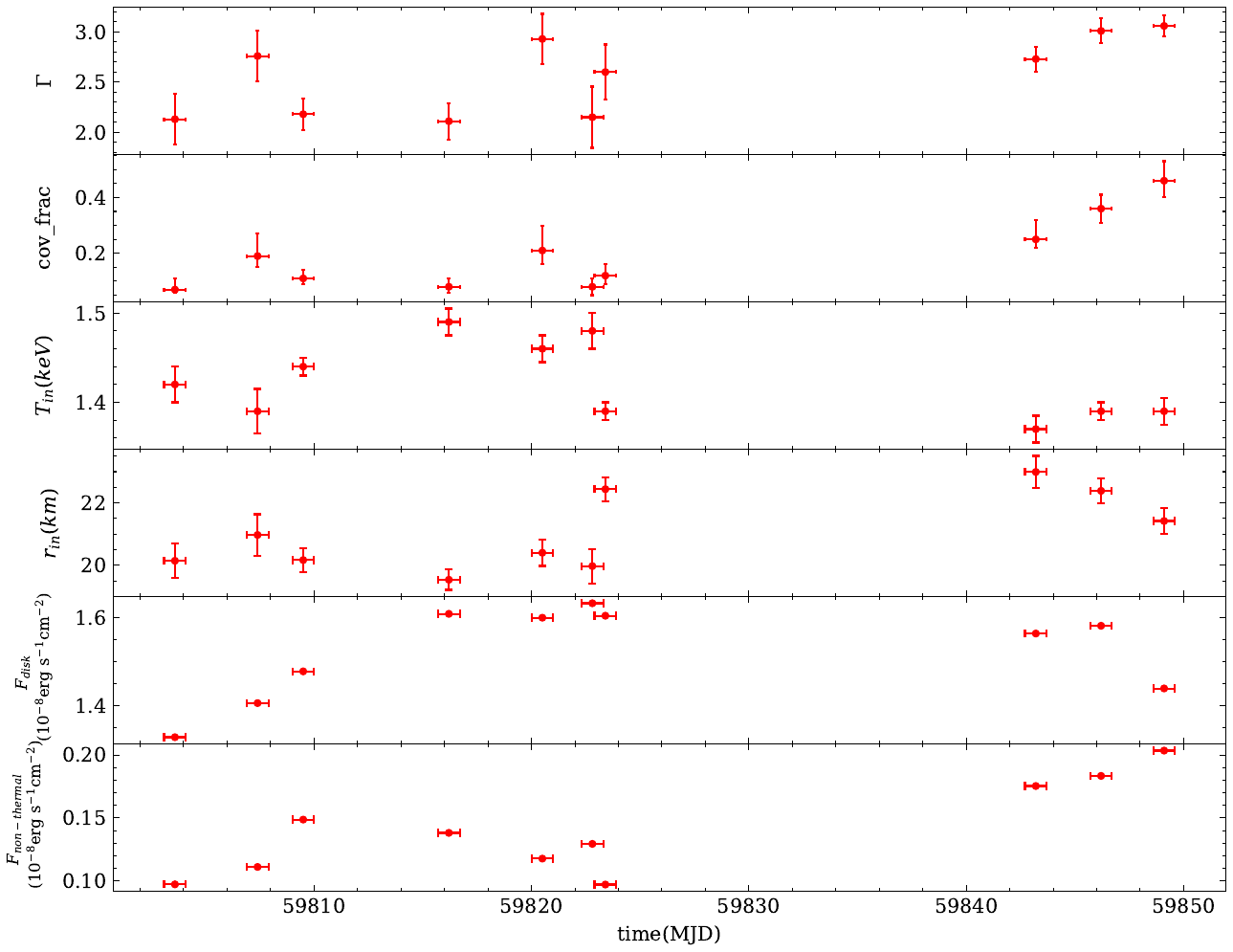}
\end{minipage}%
\hfill
\begin{minipage}[t]{0.45\linewidth}
\includegraphics[angle=0,scale=0.4]{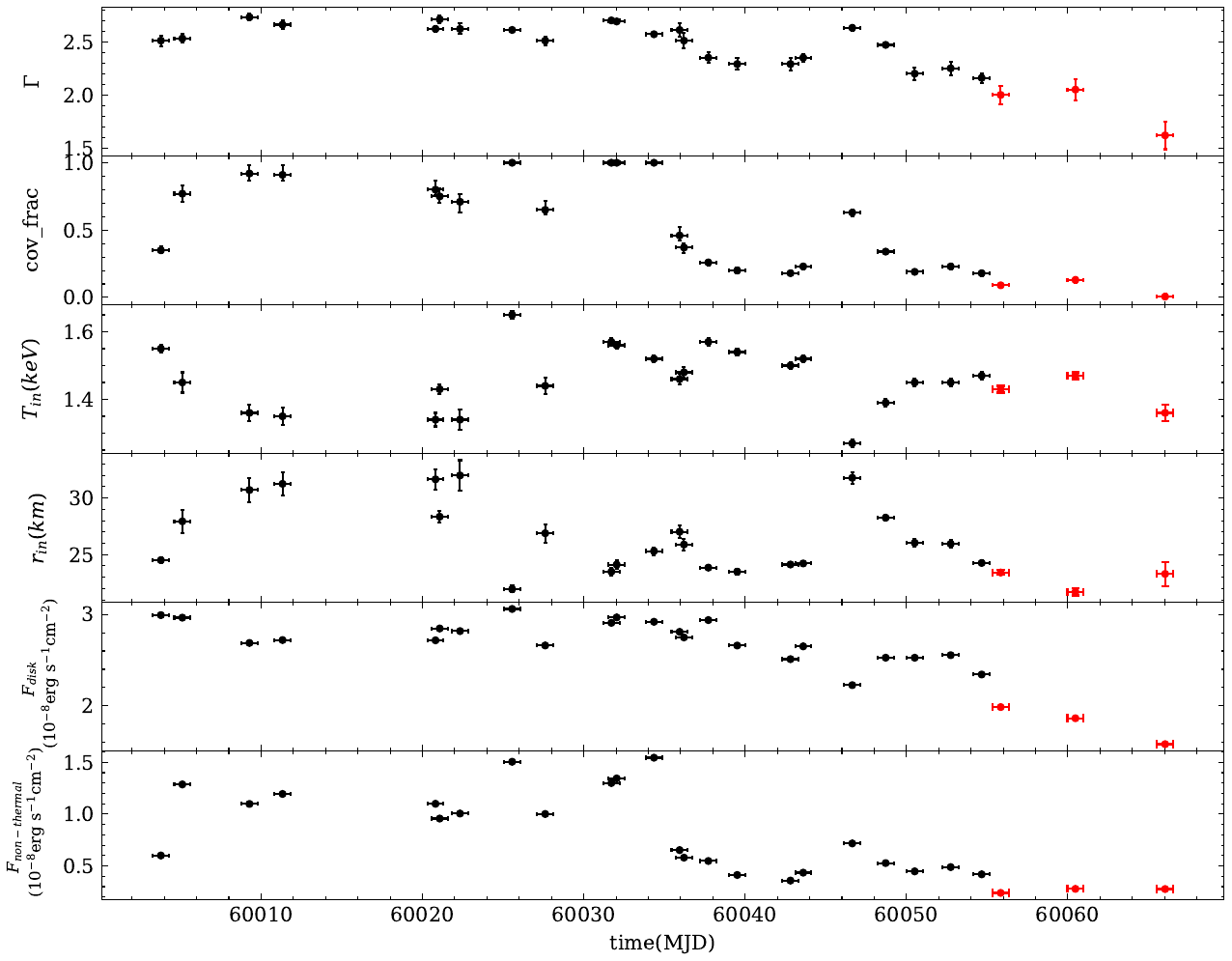}
\end{minipage}
        \caption{ Evolution of the spectral parameters of 4U 1630--472 in 2022 and 2023 from model M1: $\Gamma$ is the low-energy power-law photon index,$T_{\rm in}$ the temperature of the inner disk,  Cov\_frac the coverage factor,  and $r_{\rm in}$ the inner radius of the disk.  $F_{\rm disk}$ and $F_{\rm non-thermal}$ are disk flux and non-thermal flux, respectively.
        The red dots represent data points with relatively low fluxes, including the 2022 outburst and three low-flux data in the 2023 outburst, while the black dots correspond to high flux data for 2023.}
        
        \label{thdisk}
\end{figure*}

\begin{figure}
	\centering
	\includegraphics[angle=0,scale=0.4]{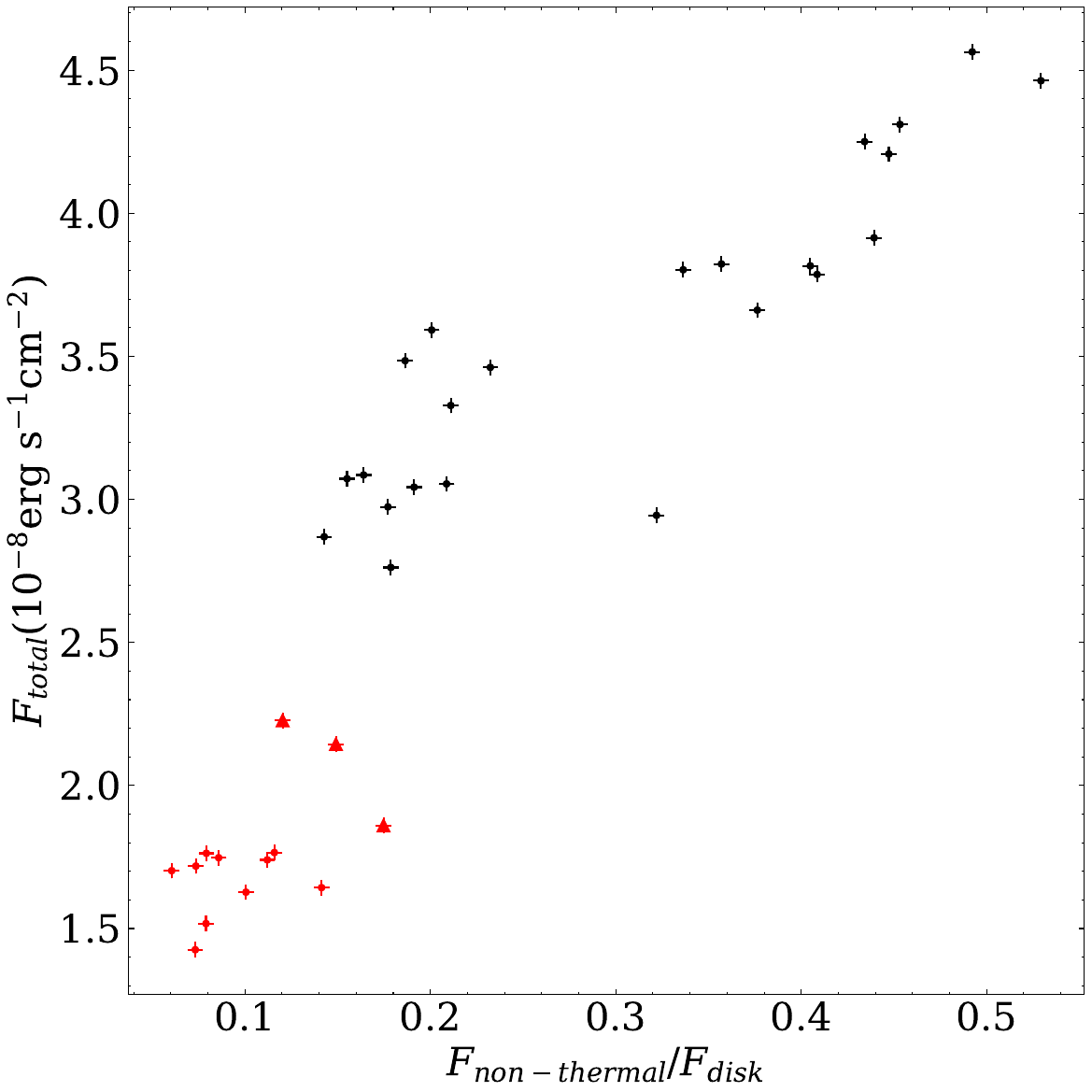}
	\caption{ The hardness-intensity diagram of 4U 1630--472, where the hardness is defined as the ratio of $F_{\rm disk}$to $F_{\rm non-thermal}$.  The colors inherit those in Figure \ref{thdisk}. The three data with the lowest flux in the 2023 outburst are plotted with triangles.  }
	\label{hidFLUX}
\end{figure}

\begin{figure}
        \centering
        \includegraphics[angle=0,scale=0.3]{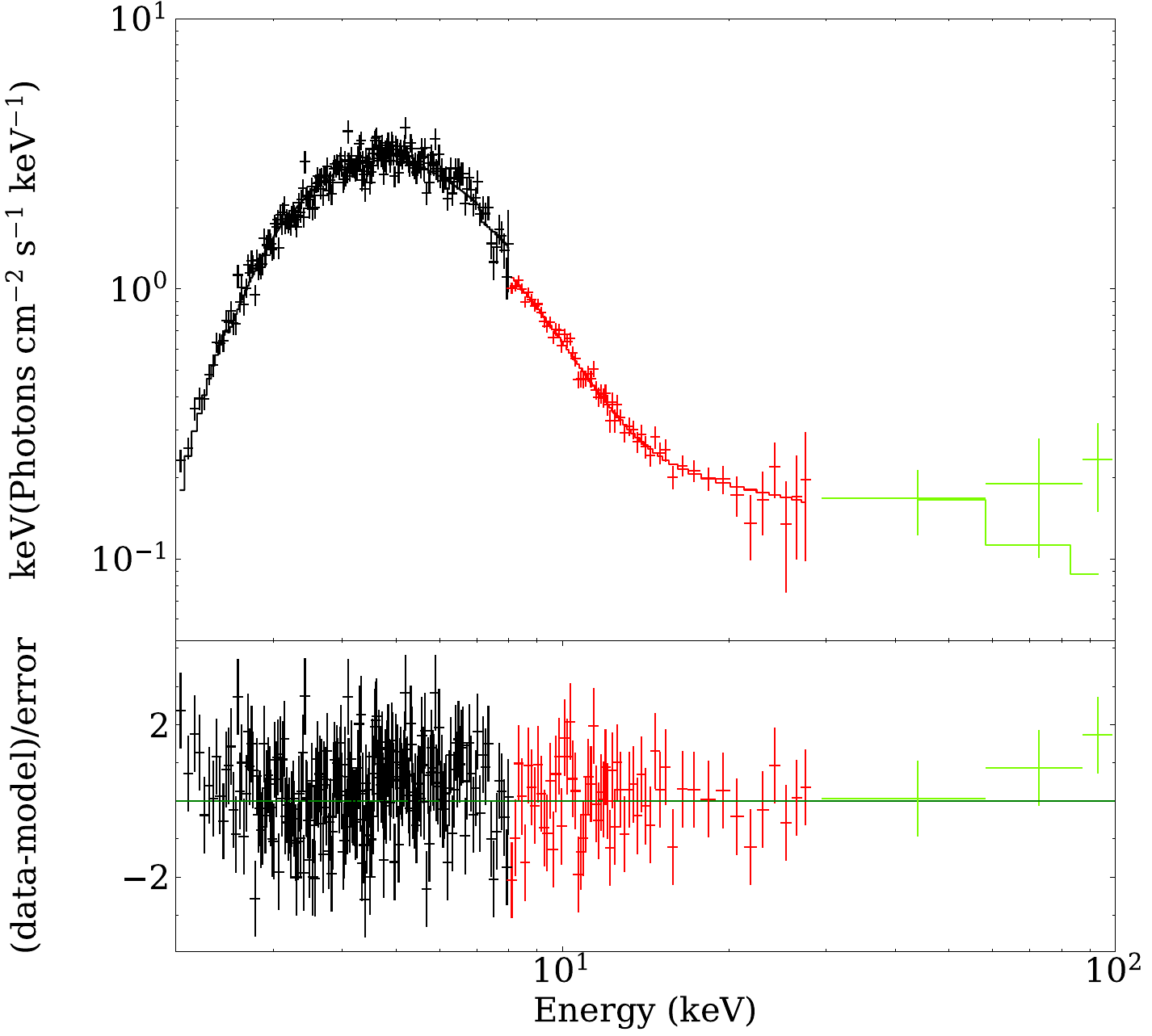}
        \caption{ Spectral fit of M1 for 4U 1630--472 of Obs P040426302301  The black, red, and green symbols correspond to the LE, ME, and HE data from Insight-HXMT, respectively.
}
        
        \label{diskbbfit}
\end{figure}

\begin{figure*}
\centering

\begin{minipage}[t]{0.5\linewidth}
\centering
\includegraphics[angle=0,scale=0.4]{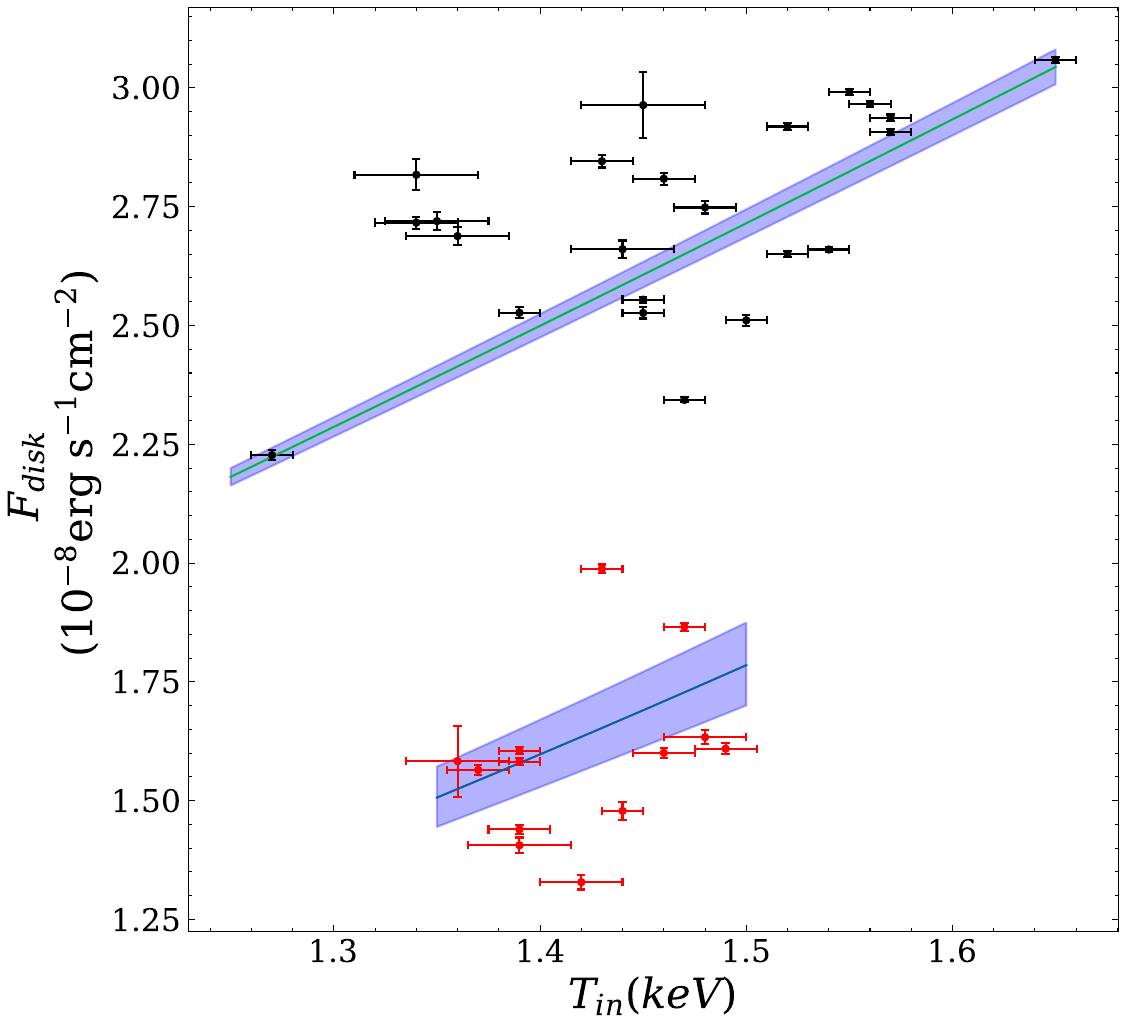}
\end{minipage}%
\hfill
\begin{minipage}[t]{0.5\linewidth}
\centering
\includegraphics[angle=0,scale=0.5]{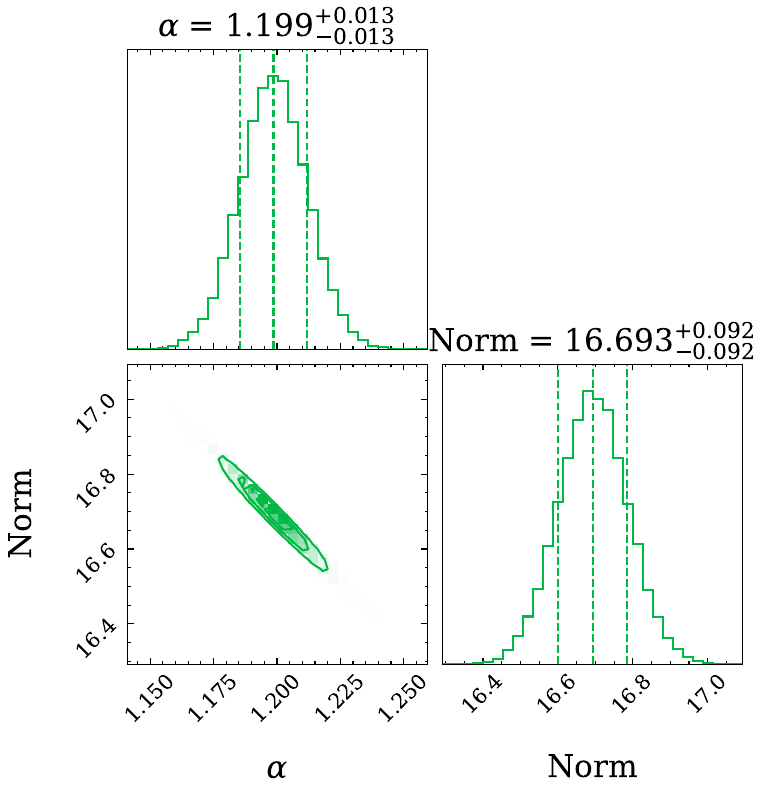}
\end{minipage}%
\hfill
\begin{minipage}[t]{0.4\linewidth}
\centering
\includegraphics[angle=0,scale=0.5]{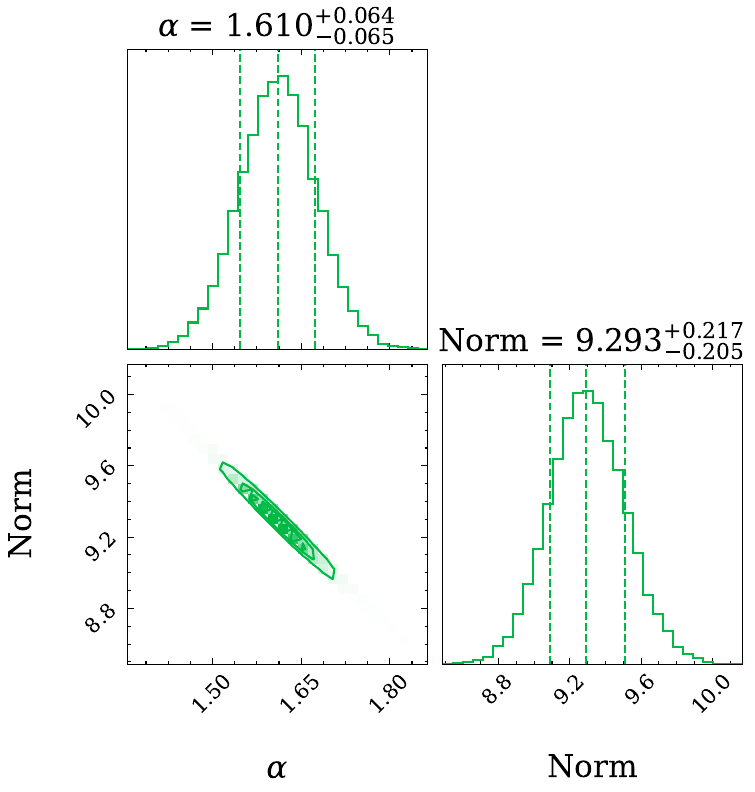}
\end{minipage}

\centering
\caption{ The relationship between inner disk temperature and flux. Red and black points inherit the colors illustrated in Figure \ref{thdisk}. The green and blue lines represent the relationship of  $F_{\rm disk} \propto T_{\rm in}^{\alpha}$ , where $\alpha_{1}$ = $1.20\pm 0.01$ and $\alpha_{2}$ = $1.61 \pm 0.06$.
The two subfigures below are One-and two-dimensional projections of the posterior probability distributions, and the 0.16, 0.5 and 0.84 quantile contours derived from the MCMC analysis for two parameters.}
        
        \label{T4}
\end{figure*}

\begin{figure*}
\centering

\begin{minipage}[t]{0.45\linewidth}
\centering
\includegraphics[angle=0,scale=0.4]{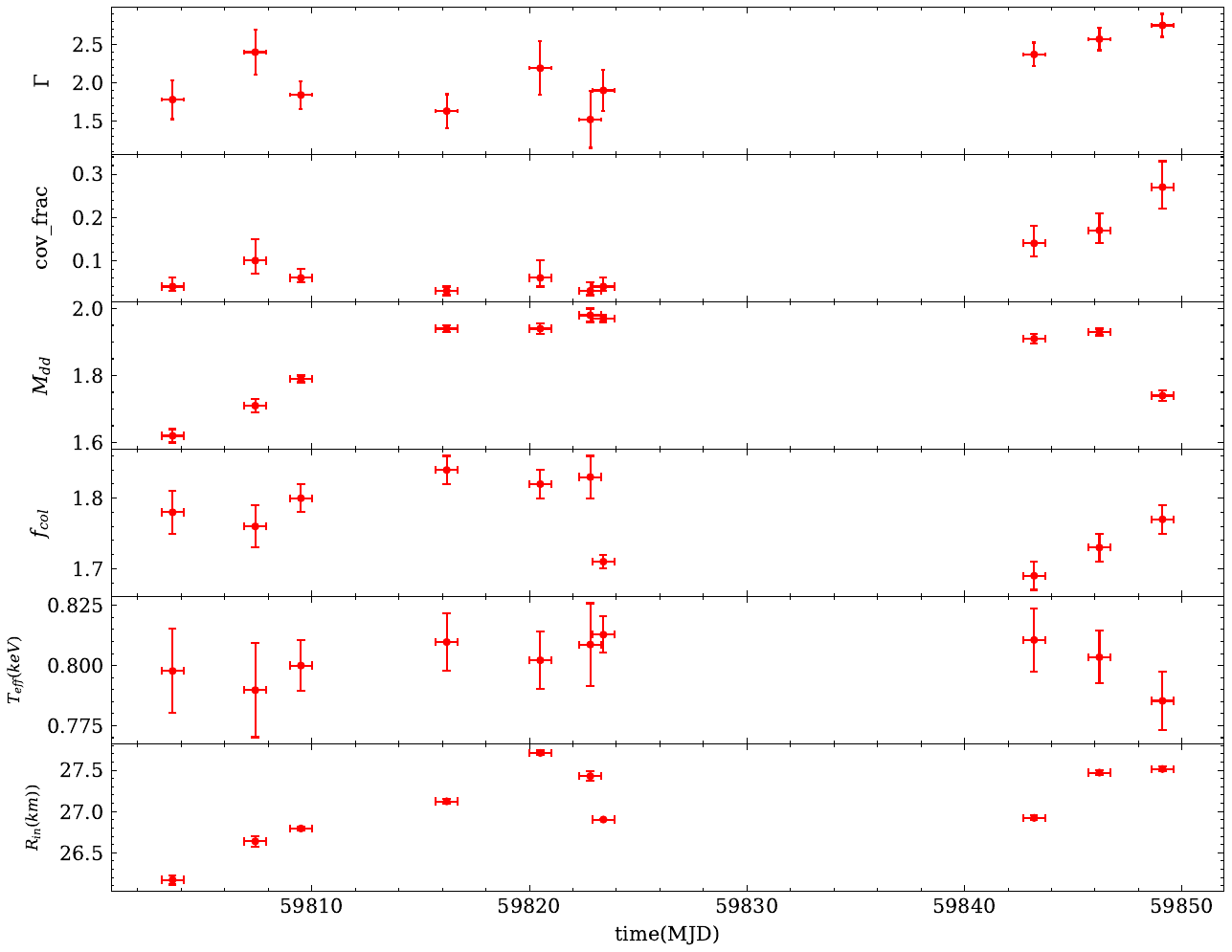}
\end{minipage}%
\hfill
\begin{minipage}[t]{0.45\linewidth}
\includegraphics[angle=0,scale=0.4]{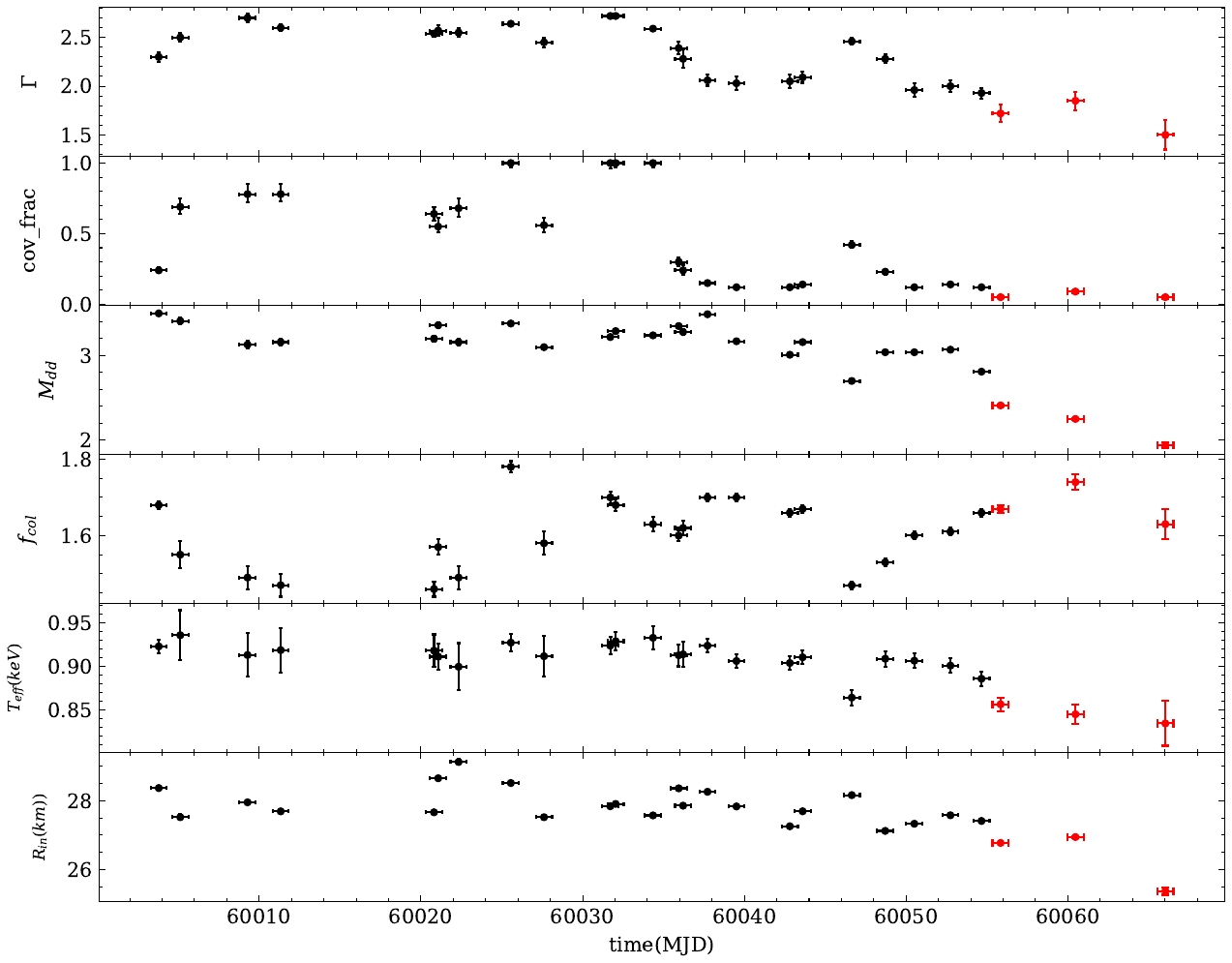}
\end{minipage}
        \caption{ Evolution of the spectral parameters of 4U 1630--472 in 2022 and 2023 from model M2: $\Gamma$ is the low-energy power-law photon index, Cov\_frac the coverage factor, $M_{\rm dd}$ the “effective” mass accretion rate of the disk in units of $10^{18}$ g/s,  $f_{\rm col}$ the spectral hardening factor and 
        $T_{\rm eff}$ the effective temperature of the inner zone of the disk, obtained by $T_{\rm in}/f_{\rm col}$, and $R_{\rm in}$ the inner radius of the disk. The colors inherit those in Figure \ref{thdisk}. }
        
        \label{kerr}
\end{figure*}

\begin{figure}
        \centering
        \includegraphics[angle=0,scale=0.3]{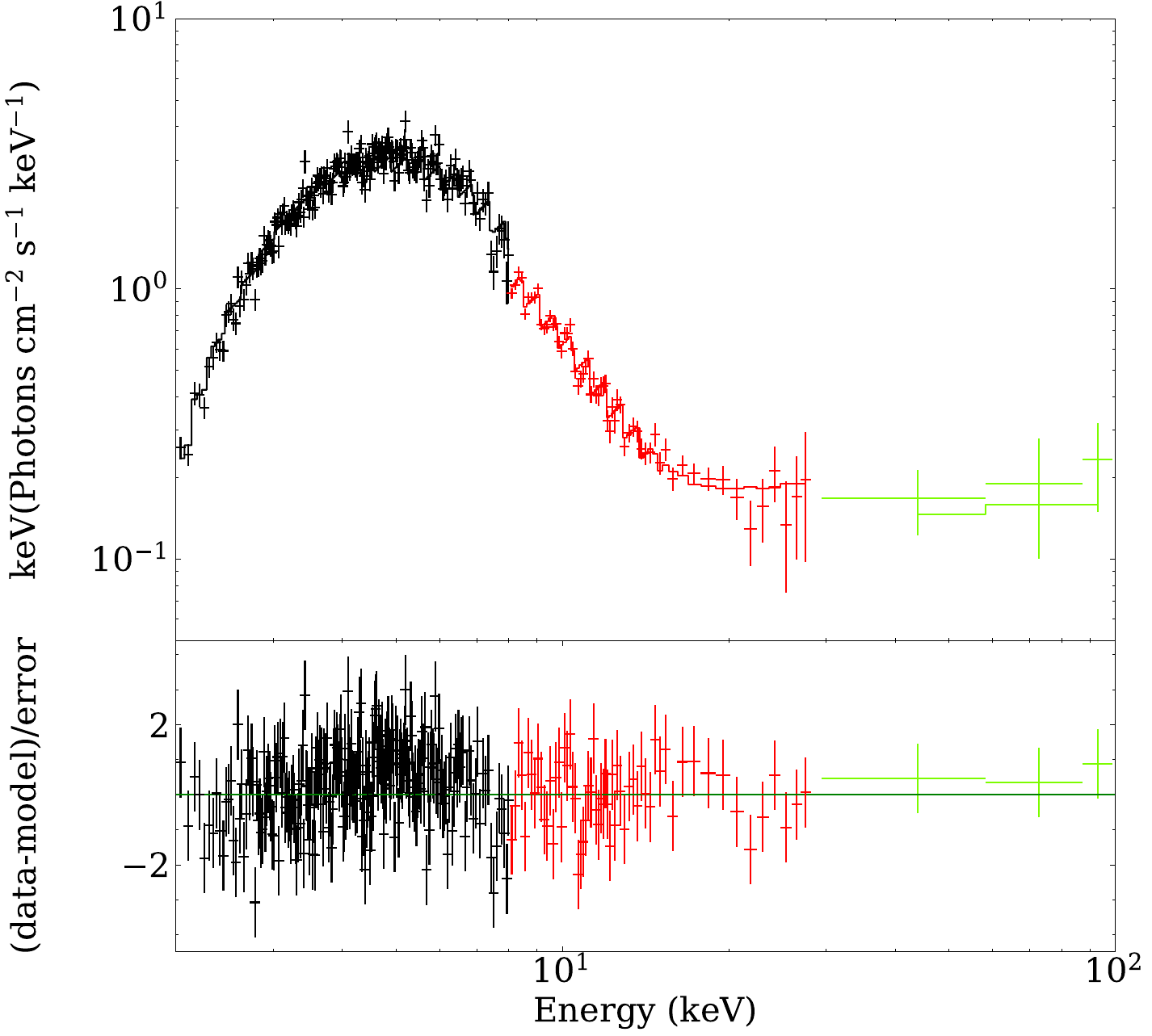}
        \caption{ Spectral fit of M2 for 4U 1630--472 of Obs P040426302301  The black, red, and green symbols correspond to the LE, ME, and HE data from Insight-HXMT, respectively.
}
        
        \label{kerrbbfit}
\end{figure}

\begin{figure*}
\centering

\begin{minipage}[t]{0.1\linewidth}
\centering
\includegraphics[angle=0,scale=1]{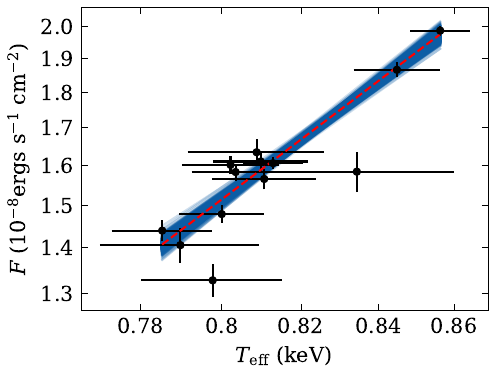}
\end{minipage}%
\hfill
\begin{minipage}[t]{0.5\linewidth}
\centering
\includegraphics[angle=0,scale=1]{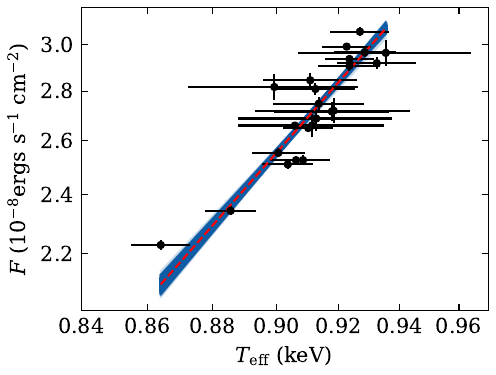}
\end{minipage}%
\hfill
\begin{minipage}[t]{0.5\linewidth}
\centering
\includegraphics[angle=0,scale=0.6]{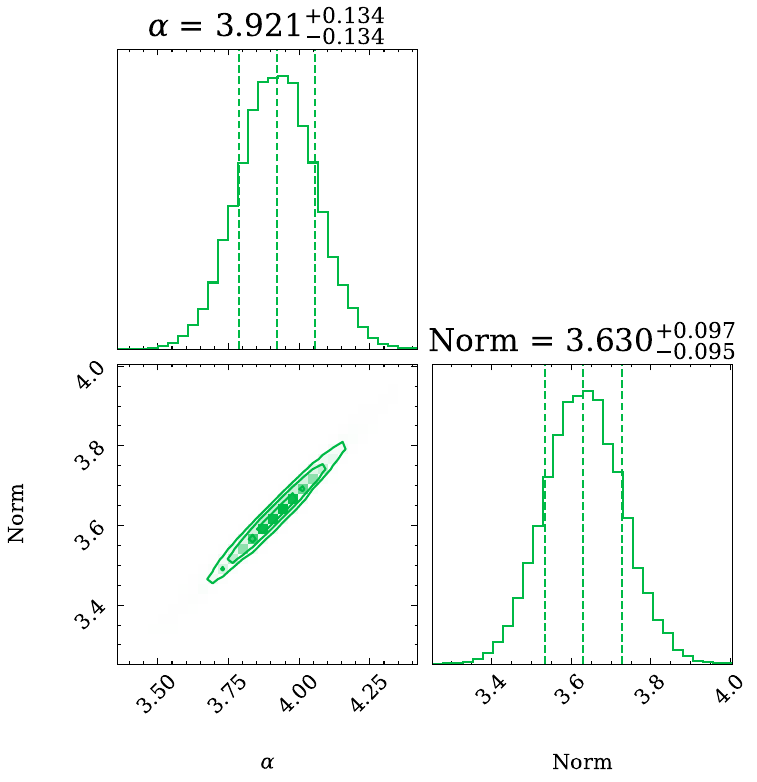}
\end{minipage}
\hfill
\begin{minipage}[t]{0.45\linewidth}
\centering
\includegraphics[angle=0,scale=0.6]{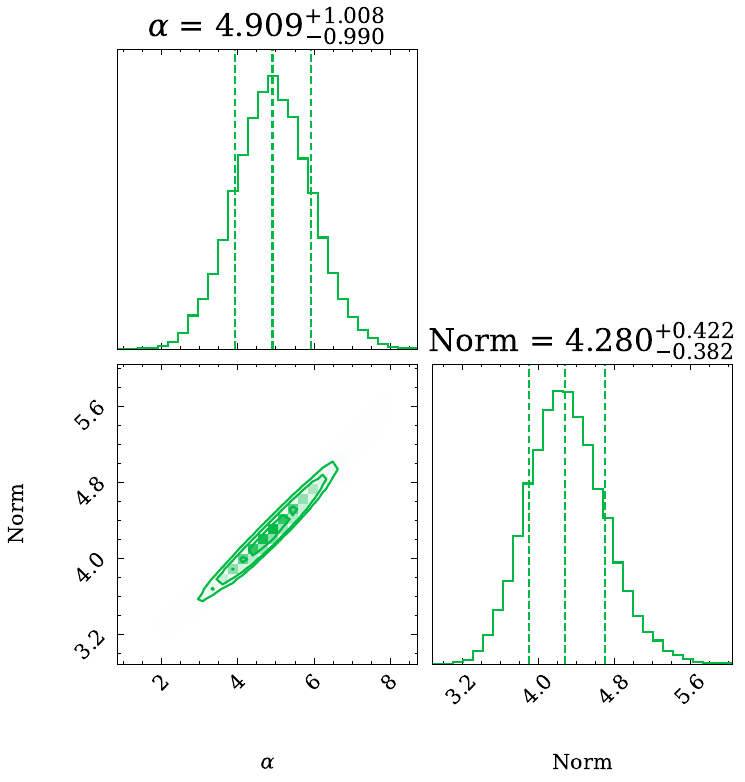}
\end{minipage}

\centering
\caption{ The disk unabsorbed flux (1--100 keV) versus the disk inner temperature ($T_{\rm eff}$). Flux correlates with temperature in forms of $T^{3.92 \pm 0.13}_{\rm eff}$ and $T^{4.91 \pm 1.00}_{\rm eff}$ during the 2022 (left) and 2023 (right) outbursts, respectively. The two subfigures below are One-and two-dimensional projections of the posterior probability distributions, and the 0.16, 0.5 and 0.84 quantile contours derived from the MCMC analysis for two parameters.}
\label{M-T}
\end{figure*}

\begin{figure}
        \centering
        \includegraphics[angle=0,scale=0.4]{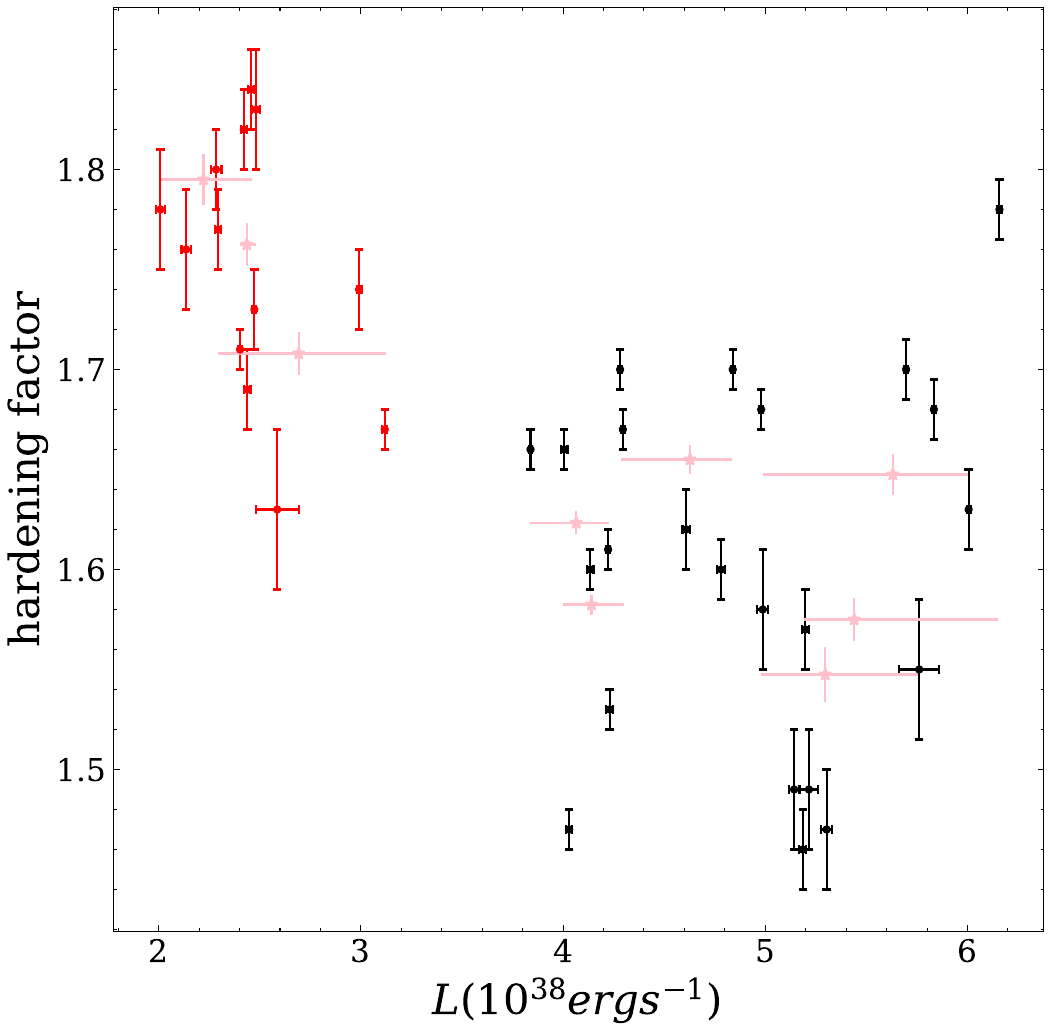}
        \caption{Evolution of hardening factor with luminosity.  The pink stars are the mean values calculated by combining the neighboring four data.}
        
        \label{hd-f}
\end{figure}

\begin{figure*}
\centering

\begin{minipage}[t]{0.45\linewidth}
\centering
\includegraphics[angle=0,scale=0.4]{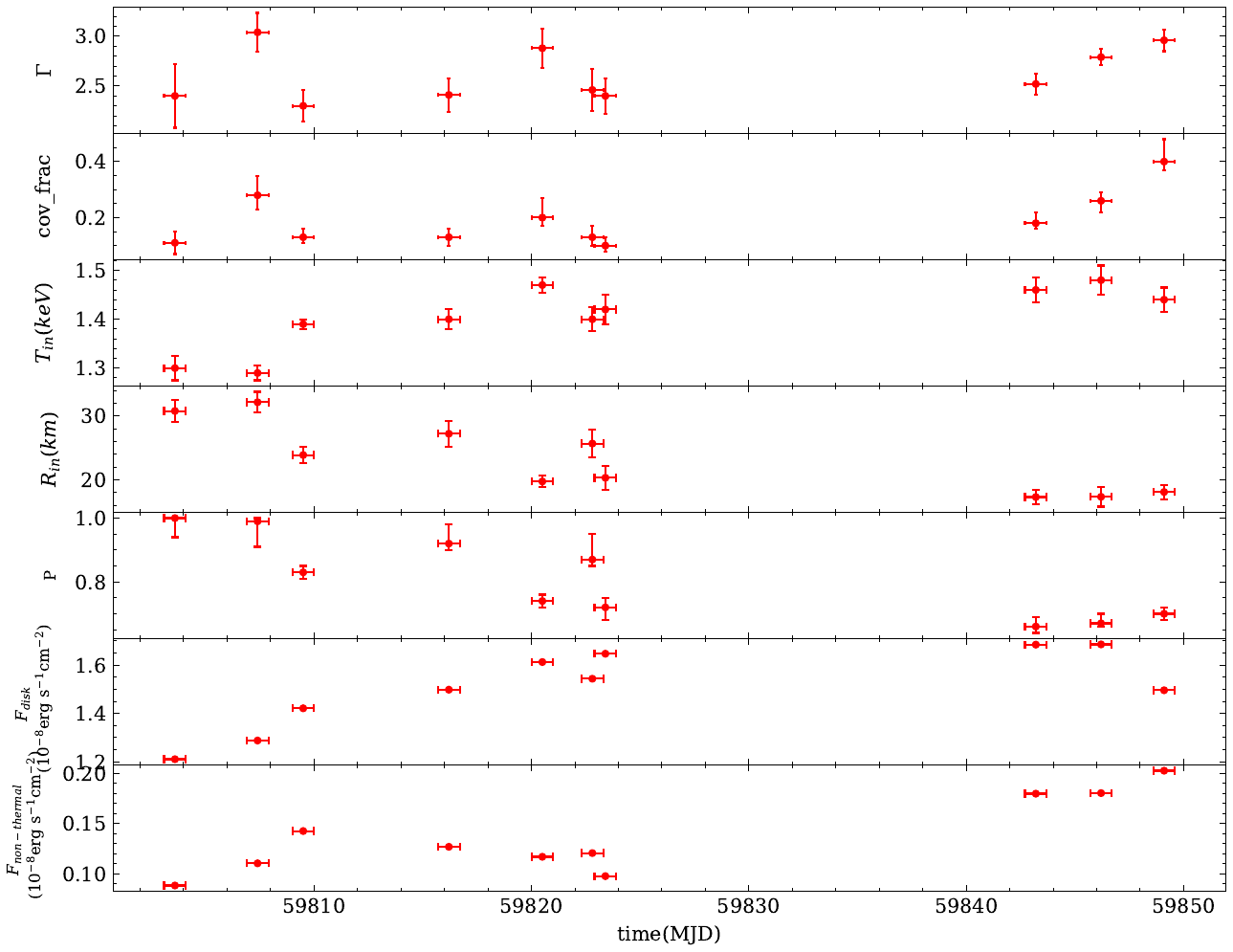}
\end{minipage}%
\hfill
\begin{minipage}[t]{0.45\linewidth}
\includegraphics[angle=0,scale=0.4]{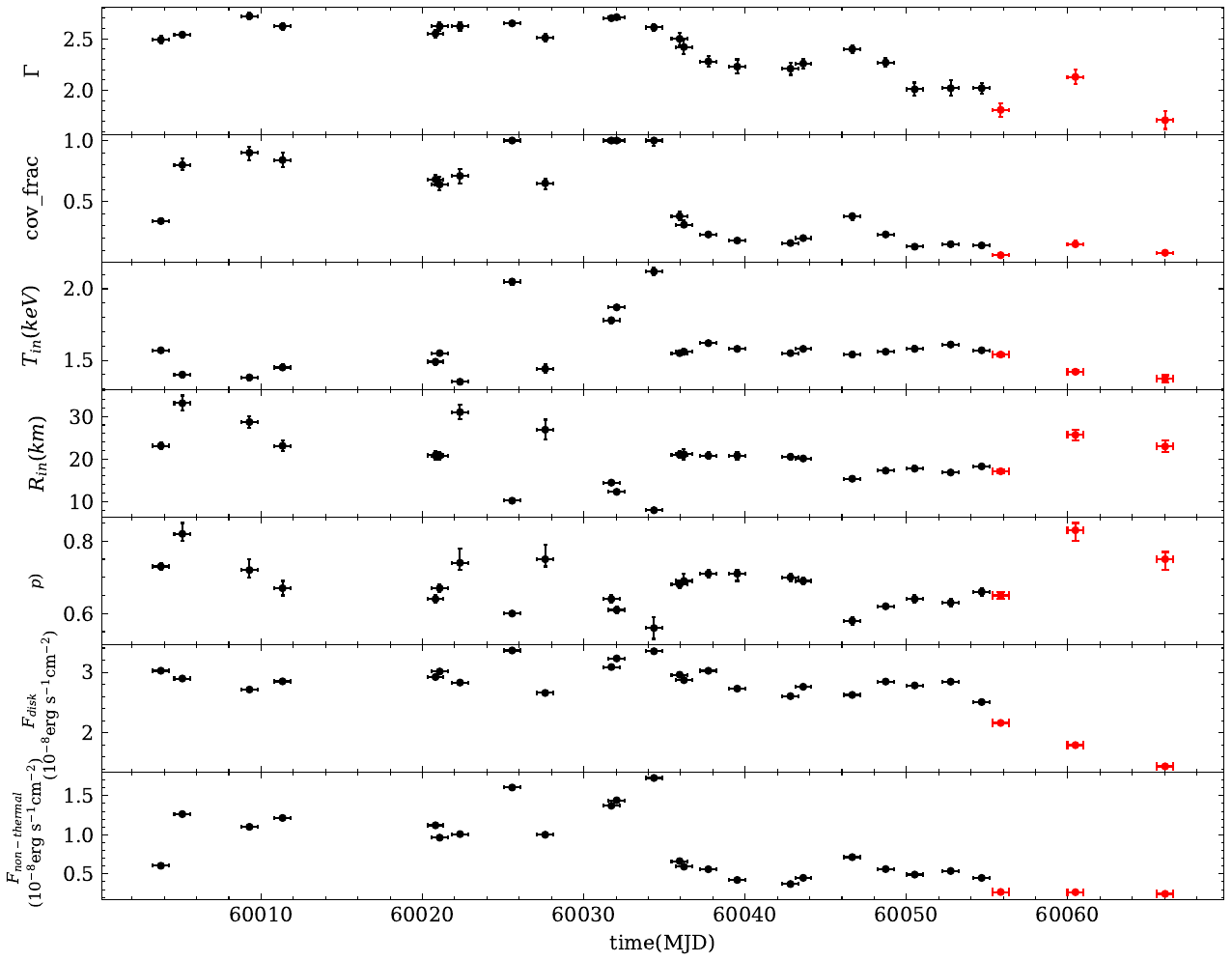}
\end{minipage}
        \caption{ Evolution of the spectral parameters of 4U 1630--472 in 2022 and 2023 from model M3: $\Gamma$ is the low-energy power-law photon index, $T_{\rm in}$ the temperature of the inner disk,  Cov\_frac the coverage factor,  and $R_{\rm in}$ the inner radius of the disk, $p$ the exponent of the radial dependence of the disk temperature, $F_{\rm disk}$ and $F_{\rm non-thermal}$ are disk flux and non-thermal flux, respectively. The colors inherit those in Figure \ref{thdisk}.}
        
        \label{diskpbb}
\end{figure*}

\begin{figure}
        \centering
        \includegraphics[angle=0,scale=0.3]{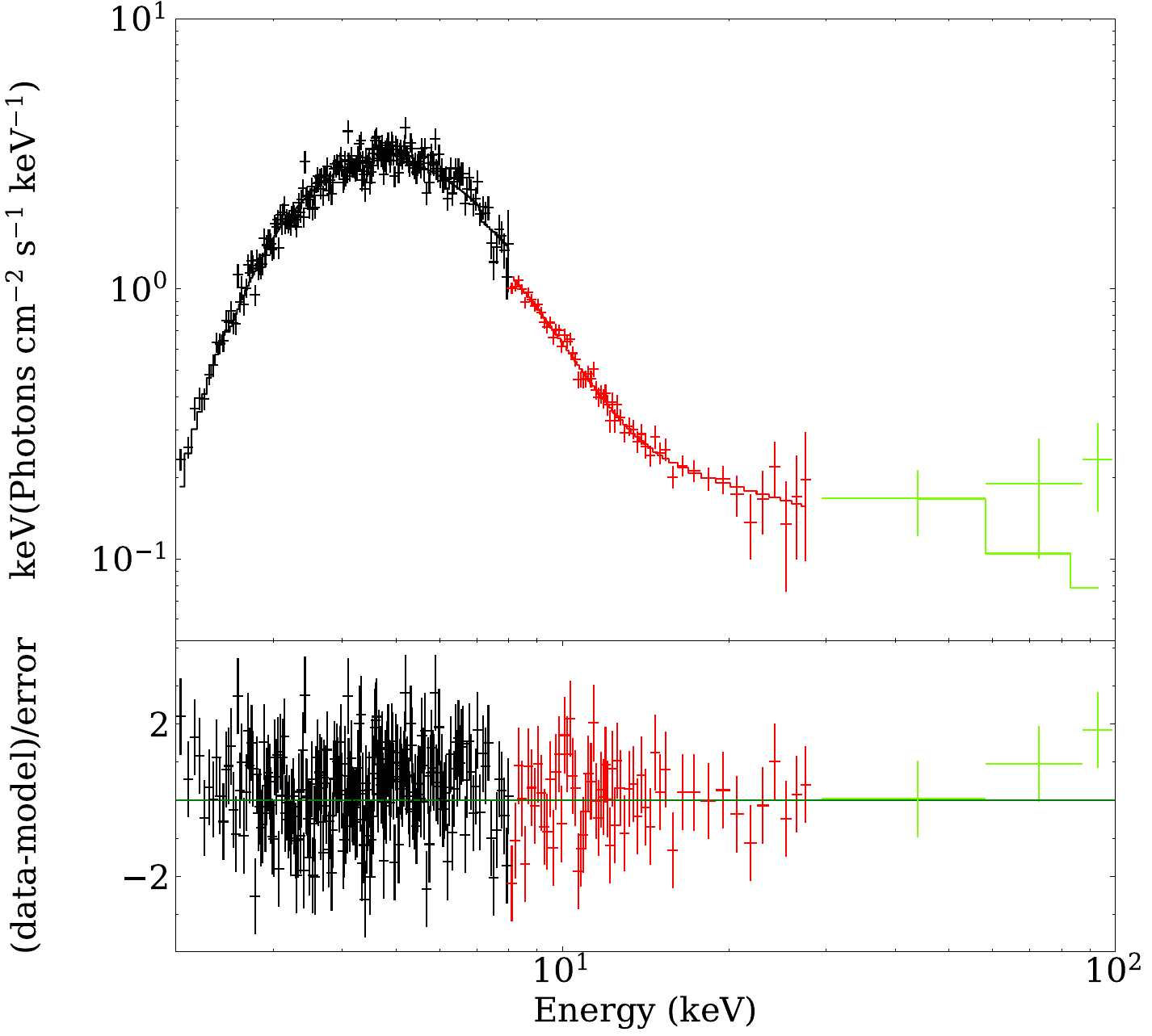}
        \caption{ Spectral fit of M3 for 4U 1630--472 of Obs P040426302301  The black, red, and green symbols correspond to the LE, ME, and HE data from Insight-HXMT, respectively.}
        
        \label{diskpbbfit}
\end{figure}

\begin{figure}
        \centering
        \includegraphics[angle=0,scale=0.4]{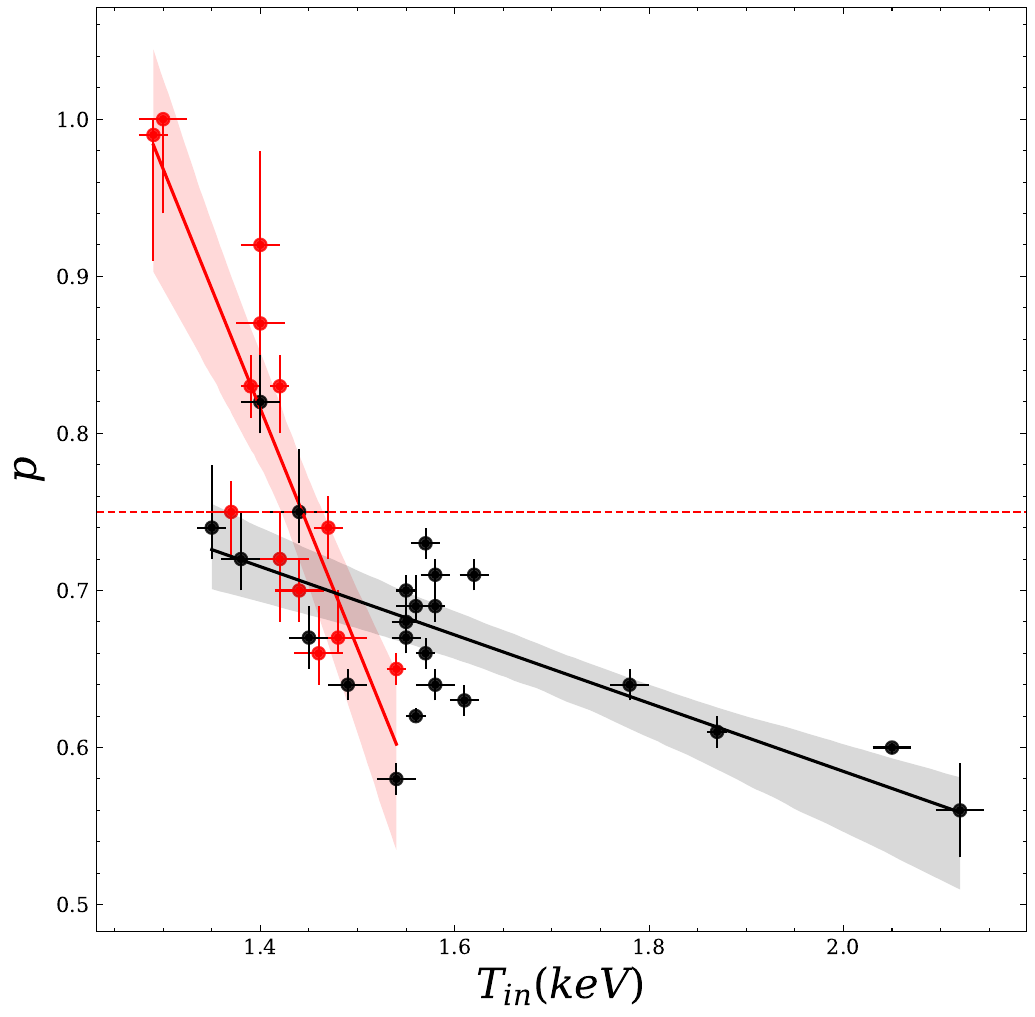}
        \caption{ Evolution of $p$ against inner disk temperature $T_{\rm in}$. Colors inherit those in Figure \ref{thdisk}.  The grey lines are the fits with linear slopes of -1.53 and -0.22.  Shaded areas are 95\% confidence intervals.}
        
        \label{ptin}
\end{figure}

The spectral fitting is carried out using the software package {\tt{XSPEC V12.13.1}}. We adopt several models to fit the spectrum of Insight-HXMT. 
The {\tt tbabs} is taken to account for interstellar absorption \citep{2000Wilms},  with photoelectric cross-sections provided by \cite{1996Verner}, and the disk component is fitted with  {\tt diskbb} \citep{1984Mitsuda}.
To consider the differences in the effective area of the calibration between different instruments \citep{2018Chen}. 
{\tt Constant} is introduced to balance the calibration discrepancies of Insight-HXMT LE, ME, and HE (in this paper, we fix the constant of LE to 1).  At this time, we find that the fit is not good ($\chi^2$/(d.o.f)=2067.17/897=3.25) and the residual showing up in the middle energies stimulates us to improve the fit by convolving with an additional component of {\tt thcomp}. Finally, the fit is largely improved $\chi^2$/(d.o.f)=879.29/895=0.98 (see Figure \ref{diskbbfit}). Therefore our fitting model M1 is:{\tt constant*tbabs*thcomp*diskbb}. The flux of the disk in the 1--100 keV is estimated with {\tt cflux}.  we first act cflux on diskbb to compute the thermal flux. Then act on the comp to get the total flux (thermal plus non-thermal flux) then subtract the thermal flux to get the non-thermal flux.

The parameters obtained from spectral fitting with M1 are shown in Table \ref{HXMTparadisk}  of the appendix A.
The evolutions of the spectral parameters are shown in Figure \ref{thdisk}. 
The three data points in the ending phase of the 2023 outburst exhibited lower fluxes, leading us to group them with the 2022 outburst data (represented by the red points in Figure \ref{thdisk}).  As shown in Figure \ref{hidFLUX}, from the HID with the
hardness ratio defined as the ratio of the non-thermal flux to the disk flux,
it seems  that these three data points (plotted in triangles in Figure \ref{hidFLUX}) are more aligned with the 2022 outburst in terms of their position on the HID.
We conduct a KS-test on the red and black data in Figure \ref{thdisk} and find that their p-values are 0.0002, indicating that they are not suggestive of the same distribution. We also conducted a KS-test on the last three red data and the data from the 2022 outburst, with a p-value of 0.14, indicating that they are likely suggestive of the same distribution.
We free the column density $N_{H}$ during the fitting of the outburst in 2022 and 2023 and get values around 6.5$\times 10^{22}$ $\rm cm^{-2}$, and thus we fix it at 6.5$\times 10^{22}$ $\rm cm^{-2}$ thereafter.
The inner disk radius $r_{\rm in}$ can be estimated from the normalization of the disk obtained from M1:  $\text{norm} = \left(\frac{\text{rin}}{\text{D}_{10}}\right)^2 \cos\theta$, where $r_{\rm in}$ is the inner disk radius, $D_{\rm 10}$ is the distance of the source in units of 10 kpc (10 kpc), and $\theta$ is the inclination angle (65\textdegree). 

As shown in Figure \ref{T4}, 
We employ the Markov Chain Monte Carlo (MCMC) method with the Goodman-Weare algorithm employing 8 walkers and a total chain length of 40,000 iterations \citep{2010G}. The initial 2000 elements of the chain were discarded as a "burn-in" period to ensure convergence and stability of the fit.
The distributions of the inner disk temperature ($T_{\rm in}$) and the flux of the disk ($F_{\rm disk}$) are diffuse and power-law fits give  $F_{\rm disk} \propto T_{\rm in}^{1.20 \pm 0.01}$ and $F_{\rm disk} \propto T_{\rm in}^{1.61 \pm 0.06}$, which are largely deviating from $T_{\rm in}^{4}$ expected for a standard disk. Such deviation could potentially be attributed to the variable hardening factor. Consequently, our measured  $r_{\rm in}$ and $T_{\rm in}$ may not represent the intrinsic ones.

\subsubsection{ kerrbb}

To investigate the effect of the hardening factor on measurements of the temperature and the inner radius of the disk, we replace {\tt diskbb} with {\tt kerrbb} thus have the have model M2: {\tt constant*tbabs*thcomp*kerrbb}. With eta, mass, spin, inclination, and distance fixed at 0, 10 $ M_\odot$, 0.9, 65°, and 10 kpc, respectively \citep{1998Kuu,2014S,2018K}, and normalization at 1 as required by the model,  fit with model M2 results in $\chi^2$/(d.o.f)=1.15 (see Figure \ref{kerrbbfit}).

The spectral parameters born out of M2 are shown in Figure \ref{kerr} and Table \ref{HXMTparakerrbb}  of the appendix A., where $R_{in}$ is calculated with $R_{\rm in} =f_{col}^{2} \xi r_{\rm in}$ \citep{2000M},  here, $f_{col}$  is a spectral hardening factor \citep{1995S}, and $\xi$ is set to 0.41 \citep{1998K} to correct the inner boundary condition.
The effective temperature ($T_{\rm eff}$) is determined by $T_{\rm in}$ with the harding factor.
After introducing the hardening factor, the inner radius $R_{in}$ becomes relatively stable, and the temperature dependence of the disk flux turns out to be $T_{\rm eff}^{3.92 \pm 0.13}$ and $T_{\rm eff}^{4.91 \pm 1.00}$ for the 2022 and 2023 outbursts, respectively, with the same fitting method as in Figure \ref{T4} (Figure \ref{M-T}). 
However, it is noticeable that the uncertainties in the power law indices obtained from the fitting are relatively large.
Figure \ref{hd-f} shows the flux dependence of the hardening factor derived from the 2022 and 2023 outbursts. The pink stars are the mean value calculated by dividing the four data into one group.
We find that the hardening factor decreases with increasing luminosity, but the distribution is more diffuse at high luminosity. The Spearman rank-order correlation coefficient and p-value were obtained by the Spearman rank-order test as -0.73 and 0.024 respectively, briefly consistent with the theoretical evolution trend from \cite{2000Merloni,2022Ren}.  Given the flux-dependence of the hardening factor and the rough consistency with $T_{\rm in}^4$ of the disk flux, 2022 and 2023 outbursts may still hold a standard disk. 

\subsubsection{diskpbb}

To further investigate the possible deviation of 4U 1630--472 disks from a standard disk, we replace {\tt diskbb} with {\tt diskpbb}. Therefore, our model 3 is represented as {\tt constant*tbabs*(thcomp*diskpbb)}, and the fit results in $\chi^2$/(d.o.f.)=0.97 (see Figure \ref{diskpbbfit}). 
{\tt Diskpbb} is a multiple blackbody disk model with local disk temperature T(r) proportional to $r^{-p}$, where $p$ is a free parameter. $p$ takes 0.75 for the standard disk and 0.5 for the slim disk. \citep{1994M,1995H,2000W,2004K}.
The average of $p$ is $0.81_{-0.12}^{+0.13}$ for the 2022 outburst and $0.67_{-0.07}^{+0.09}$ for the 2023 outburst (Figure \ref{diskpbb}). For both outbursts, a standard disk can not be excluded due to that 0.75 is covered by the averaged $p$ within 1 $\sigma$ error bars, but a slim disk is not likely supported since  0.5 is not covered by the averaged $p$ even within 2 $\sigma$ of the errorbars.  The parameters obtained from spectral fitting with M3 are shown in Table \ref{HXMTparadiskpbb}  of the appendix A.
The evolution of $p$ with $T_{\rm in}$ as shown in Figure \ref{ptin}, indicates that $p$ of the 2022 outburst is generally larger than that of the 2023 outburst.  
We use the linregress function to fit the temperature and $p$. Linear fits result in slopes of -1.53 and -0.22 for the 2022 and 2023 outbursts. The reduced chi-square and degrees of freedom are 2.71, 11 and 4.33, 21, respectively.
We also test their correlation using Spearman rank-order and obtain correlation coefficients and p-values of -0.86, 0.0002 and -0.56, 0.005 respectively.
It turns out that, for outbursts from 2022 to 2023, $p$ decreases monotonously with temperature and thus suggests a disk may continuously evolve toward a slim one with decreasing radiative efficiency.

\section{discussion and conclusion}
\label{dis}

\begin{figure}
        \centering
        \includegraphics[angle=0,scale=1]{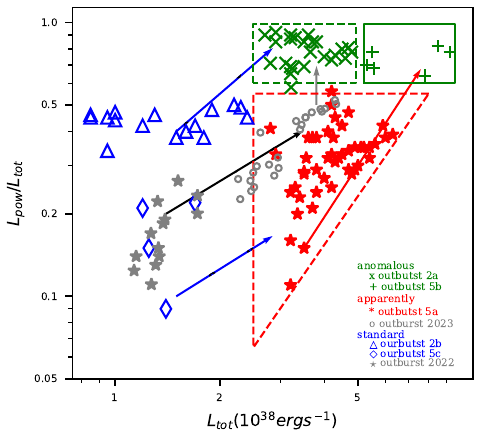}
        \caption{ Luminosity ratio of the non-thermal to the total obtained with model M1,  plotted against the total luminosity.
        The green rectangle and red dashed triangle represent the anomalous and apparently standard regimes defined in \cite{2005A}.  The RXTE data with blue triangles and diamonds are relevant to the standard regime  \cite{2005A}. 
        The blue arrows represent the evolution with luminosity in \cite{2005A}.
        The grey color represents the data from our article, with the star symbol representing the outburst in 2022 and the dot symbol the outburst in 2023. The black arrow indicates the direction of luminosity evolution, while the grey arrow shows the possible evolving direction at higher luminosities.  }
        
        \label{ratio}
\end{figure}

We have conducted spectral analyses of the outbursts of 4U 1630--472 observed by  Insight-HXMT in 2022 and 2023.  We find that during the 2022 and 2023 outbursts, the distributions of the inner disk temperature against the flux are scattered around $F_{\rm disk} \propto T_{\rm in}^{1.2}$ and $F_{\rm disk} \propto T_{\rm in}^{1.6}$, which are not consistent with the standard disk hypothesis. By considering additionally the hardening factor, these relations move to $F \propto T_{\rm eff}^{3.92\pm 0.13}$ and $F \propto T_{\rm eff}^{4.91\pm 1.00}$. With a $p$-free model, although the standard disk can not be excluded,  an evolution of decreasing $p$ with disk temperature is visible. 

\cite{2005A} classified the soft state of 4U 1630--472 into three states with RXTE observations of five outbursts spanning from 1996 to 2004.
In the first state, the disk behaves like a standard one. The accretion disk model can well explain the X-ray spectrum of this state, the disk temperature $T_{\rm in}  \textless$ 1.2keV, and the X-ray luminosity is less than $2.5 \times10^{38}erg/s$. The disk flux follows the $L_{\rm disk} \propto T_{\rm in}^4$, indicating that the accretion disk structure is consistent with the standard thin disk model.
In the second state,
the spectrum appears similar to the standard state, but the relationship between disk flux and temperature is $L_{\rm disk} \propto T_{\rm in}^2$, indicating that the accretion disk radiation efficiency is relatively low. An optically thick, convection-dominated "slim disk" may have formed.
In the third state, the accretion disk temperature and luminosity are similar to the standard state, but there is a strong high-energy Comptonized component. This may be due to the inverse Compton scattering process converting part of the disk radiation into a high-energy component.  
Accordingly, they defined three regimes in the hard emission fraction-total luminosity diagram: standard, apparently standard, and anomalous regimes. 
The outbursts in 2022 and 2023 as observed by Insight-HXMT contribute two more samples for investigating the evolution of the accretion disk between different regimes.

For the 2022 and 2023 outbursts,  model M1 gives the disk flux-temperature relations not following  $F \propto T_{\rm in}^{4}$.
Scattering by the accretion disk atmosphere can lead to disk emission deviating from a standard correlation with the temperature \citep{1974M}. 
\cite{1995Shimura} demonstrated that such a deviation can be largely alleviated by introducing a hardening factor. 
After considering the impact of the hardening factor on the temperature and radius of the disk, the flux and temperature of the disk can be presented in functions of  $F \propto T_{\rm eff}^{3.92\pm 0.13}$ and $F \propto T_{\rm eff}^{4.91\pm 1.00}$, suggesting a standard disk may still hold in the 2022 and 2023 outbursts.
 \cite{2024R} analyzed the 2022 IXPE observations of 4U 1630--472 and concluded that the observed polarization properties are compatible with the accretion of matter onto the black hole at a mildly relativistic rate through a thin disk covered by a partially ionized atmosphere. The apparent deviation for the relation of disk flux and temperature from standard disk and the introduction of the hardening factor in outbursts of 2022 and 2023 are in line with the picture proposed with IXPE observations.

The further investigation of the disk properties with a p-free model gives averaged $p$ of $0.81^{+0.13}_{-0.12}$ for the 2022 outburst and $0.67_{-0.07}^{+0.09}$ for 2023 outburst. Again, both cover 0.75 for a standard disk but away from 0.5 for a slim disk \citep{2000PWatarai}.  
Albeit all the aforementioned disk features for the two outbursts can be roughly in line with a standard one,  $p$ is found to evolve significantly along with the outbursts. As shown in Figure \ref{ptin},  $p$ decreases monotonously with disk temperature. This suggests that the disk evolves from 2022 to 2023 toward a slim one with a decreasing radiation efficiency.  
\cite{2023MNRAS.525..661R} found that the polarization fraction in the 2023 outburst is lower than that of the 2022 outburst, and thus suggested that the decrease in the polarization fraction could be attributed to the presence of the corona. This is consistent with our finding that the non-thermal emission in the 2023 outburst occupies a fraction larger than that in the 2022 outburst.

 \cite{2005A} found from a series of outbursts of 4U 1630--472 that, the source can evolve between regimes in the diagram of non-thermal emission fraction against total luminosity. Here two more outbursts in 2022 and 2023 are added to this diagram for further tracing the disk evolution patterns. As shown in Figure \ref{ratio}, the grey dots represent the data points in this article to illustrate the evolution of the luminosity ratio (the non-thermal component to the total) against the total luminosity. The others outburst 5 and outburst 2 defined in \cite{2005A} observed by RXTE. 

 Figure \ref{ratio} shows that outbursts trace different evolutional tracks.
 For outburst 5, the source evolves from HSS into an apparently standard regime and then into the anomalous regime, along with increases in the luminosity and non-thermal fraction. It seems that 4U 1630--472 can stay at HSS with different non-thermal fractions. For example, outburst 2 has an average of the non-thermal fraction of around 0.5.  It is interesting that in outburst 2 the source evolves almost directly from HSS into the anomalous regime, with less evidence of observing in between an apparently standard pattern (i.e. likely a slim disk or a disk deviating from the standard one).  The 2022 and 2023 outbursts are enclosed in Figure \ref{ratio} and occupy the regimes roughly between outbursts 2 and 5.  It looks like that, according to \cite{2005A},  the 2022 outburst stays in the regime relevant to the standard disk 
 and the 2023 outburst enters into the apparently standard regime. 
 We test our data and \cite{2005A} data using the KS-test for the standard and apparently regimes. For the standard regime, the KS statistic (d-value) and p-value obtained are 0.17 and 0.98 which is suggestive of the same distribution, while for the apparent regime, the KS statistic and p-value obtained are 0.25 and 0.04 which is not suggestive of the same distribution.
 We find that the $p$ value in the 2023 outburst is larger than 0.5 required for a slim disk and it evolves significantly in the 2022 and 2023 outbursts in a manner decreasing with luminosity.  These findings may suggest that the previously reported apparently standard regime may not be fully relevant to a slim disk but stands for an evolution toward the slim disk.  The formation of a slim disk may be influenced, apart from the accretion rate, as well by the disk/corona properties  (e.g. non-thermal fraction) in HSS.  A slim disk may be less likely to form with the presence of the dominant non-thermal emission in HSS. The existence of a hot corona may help to ionize the outer part of the disk in the vertical direction and hence hinder the formation of a slim disk by alleviating the optical depth.

\begin{acknowledgments}
This work is supported by the National Key R\&D Program of China (2021YFA0718500), the National Natural Science Foundation of China under grants No. 12333007, U1838202,  U2038101, U1938103, 12273030, U1938107, 12027803 and 12173103.
This work made use of data and software from the Insight-HXMT mission, a project funded by the China National Space Administration (CNSA) and the Chinese Academy of Sciences(CAS). This work was partially supported by the International Partnership Program of the Chinese Academy of Sciences (Grant No.113111KYSB20190020).
This research has made use of software provided by data obtained from the High Energy Astrophysics Science Archive Research Center (HEASARC), provided by NASA’s Goddard Space Flight Center.
L. D. Kong is grateful for the financial support provided by the Sino-German (CSC-DAAD) Postdoc Scholarship Program (91839752).
\end{acknowledgments}

\appendix
\section{The spectral parameter}

\begin{table*}[]
    \centering
		\caption{Fitting results of Insight-HXMT for Model M1. $\Gamma$ is the low-energy power-law photon index, $T_{\rm in}$ the temperature of the inner disk,  Cov\_frac the coverage factor,  and $r_{\rm in}$ the inner radius of the disk.  $F_{\rm disk}$ and $F_{\rm non-thermal}$ are disk flux and non-thermal flux, respectively. }
    \resizebox{\textwidth}{!}{%
		\begin{tabular}{ccccccccc}
		  \hline
		   \hline
          Insight-HXMT & MJD&$\Gamma$& Cov\_frac&$T_{\rm in}$ &$r_{\rm in}$ &$F_{\rm disk}$&$F_{\rm non-thermal}$&$\chi^2$/dof
         \\ObsID &&&&(keV)&(km)&($10^{-8}$~erg~s$^{-1}$~cm$^{-2}$)&$(10^{-8}$~erg~s$^{-1}$~cm$^{-2}$)&
       \\ \hline
P040426302301 & 59803.6 & $2.13_{-0.13}^{+0.37}$ & $0.07_{-0.01}^{+0.04}$ & $1.42_{-0.02}^{+0.02}$ & $20.15_{-4.59}^{+4.81}$ & $1.33_{-0.02}^{+0.02}$ & $0.10_{-0.03}^{+0.03}$ & 942/895 \\
 P040426302403 & 59807.4 & $2.76_{-0.2}^{+0.3}$ & $0.19_{-0.04}^{+0.08}$ & $1.39_{-0.03}^{+0.02}$ & $20.98_{-4.66}^{+5.81}$ & $1.41_{-0.02}^{+0.02}$ & $0.11_{-0.02}^{+0.02}$ & 937/895 \\
P040426302501 & 59809.5 & $2.18_{-0.15}^{+0.17}$ & $0.11_{-0.02}^{+0.03}$ & $1.44_{-0.01}^{+0.01}$ & $20.17_{-3.98}^{+3.89}$ & $1.48_{-0.02}^{+0.02}$ & $0.15_{-0.02}^{+0.03}$ & 931/895 \\
P040426302701 & 59816.2 & $2.11_{-0.18}^{+0.18}$ & $0.08_{-0.02}^{+0.03}$ & $1.49_{-0.02}^{+0.01}$ & $19.54_{-2.92}^{+4.10}$ & $1.61_{-0.01}^{+0.01}$ & $0.14_{-0.02}^{+0.02}$ & 893/895 \\
P040426302802 & 59820.5 & $2.93_{-0.2}^{+0.3}$ & $0.21_{-0.05}^{+0.09}$ & $1.46_{-0.02}^{+0.01}$ & $20.40_{-4.04}^{+4.27}$ & $1.60_{-0.01}^{+0.01}$ & $0.12_{-0.02}^{+0.02}$ & 872/895 \\
P040426302901 & 59822.8 & $2.15_{-0.33}^{+0.28}$ & $0.08_{-0.03}^{+0.03}$ & $1.48_{-0.01}^{+0.03}$ & $19.98_{-5.26}^{+4.04}$ & $1.63_{-0.02}^{+0.02}$ & $0.13_{-0.03}^{+0.04}$ & 828/895 \\
P040426303001 & 59823.4 & $2.60_{-0.29}^{+0.26}$ & $0.12_{-0.03}^{+0.04}$ & $1.39_{-0.01}^{+0.01}$ & $22.44_{-4.27}^{+3.95}$ & $1.61_{-0.01}^{+0.01}$ & $0.10_{-0.01}^{+0.01}$ & 969/895 \\
P051435300101 & 59843.2 & $2.73_{-0.09}^{+0.16}$ & $0.25_{-0.03}^{+0.07}$ & $1.37_{-0.01}^{+0.02}$ & $22.99_{-5.10}^{+4.61}$ & $1.57_{-0.01}^{+0.01}$ & $0.18_{-0.02}^{+0.02}$ & 908/895 \\
P051435300301 & 59846.2 & $3.01_{-0.14}^{+0.11}$ & $0.36_{-0.05}^{+0.05}$ & $1.39_{-0.01}^{+0.01}$ & $22.39_{-4.07}^{+4.29}$ & $1.58_{-0.01}^{+0.01}$ & $0.18_{-0.01}^{+0.01}$ & 1023/895 \\
P051435300401 & 59849.1 & $3.06_{-0.1}^{+0.11}$ & $0.46_{-0.06}^{+0.07}$ & $1.39_{-0.02}^{+0.01}$ & $21.42_{-3.86}^{+4.64}$ & $1.44_{-0.01}^{+0.01}$ & $0.20_{-0.01}^{+0.01}$ & 968/895 \\
P050523700101 & 60003.8 & $2.51_{-0.05}^{+0.05}$ & $0.35_{-0.02}^{+0.03}$ & $1.55_{-0.01}^{+0.01}$ & $24.51_{-3.67}^{+3.77}$ & $2.99_{-0.01}^{+0.01}$ & $0.60_{-0.01}^{+0.01}$ & 810/895 \\
P050523700201 & 60005.1 & $2.53_{-0.04}^{+0.04}$ & $0.77_{-0.06}^{+0.06}$ & $1.45_{-0.03}^{+0.03}$ & $27.94_{-7.36}^{+7.84}$ & $2.96_{-0.07}^{+0.07}$ & $1.29_{-0.08}^{+0.08}$ & 926/895 \\
P050523700402 & 60009.3 & $2.73_{-0.03}^{+0.03}$ & $0.92_{-0.05}^{+0.06}$ & $1.36_{-0.02}^{+0.03}$ & $30.71_{-8.29}^{+8.05}$ & $2.69_{-0.02}^{+0.02}$ & $1.10_{-0.03}^{+0.03}$ & 833/895 \\
P050523700501 & 60011.3 & $2.66_{-0.04}^{+0.04}$ & $0.91_{-0.04}^{+0.07}$ & $1.35_{-0.03}^{+0.02}$ & $31.25_{-7.06}^{+8.82}$ & $2.72_{-0.02}^{+0.02}$ & $1.19_{-0.03}^{+0.03}$ & 859/895 \\
P050523700901 & 60020.8 & $2.62_{-0.03}^{+0.03}$ & $0.80_{-0.04}^{+0.07}$ & $1.34_{-0.02}^{+0.02}$ & $31.65_{-6.84}^{+8.31}$ & $2.72_{-0.01}^{+0.01}$ & $1.10_{-0.02}^{+0.02}$ & 804/895 \\
P050523700903 & 60021.1 & $2.71_{-0.04}^{+0.04}$ & $0.75_{-0.05}^{+0.06}$ & $1.43_{-0.02}^{+0.01}$ & $28.35_{-5.61}^{+5.50}$ & $2.85_{-0.01}^{+0.01}$ & $0.96_{-0.02}^{+0.02}$ & 866/895 \\
P050523701102 & 60022.3 & $2.62_{-0.06}^{+0.04}$ & $0.71_{-0.08}^{+0.06}$ & $1.34_{-0.02}^{+0.04}$ & $32.00_{-10.08}^{+8.39}$ & $2.82_{-0.03}^{+0.03}$ & $1.01_{-0.05}^{+0.05}$ & 935/895 \\
P050523701302 & 60025.6 & $2.61_{-0.01}^{+0.01}$ & $1.00_{-0.01}^{+0.0}$ & $1.65_{-0.01}^{+0.01}$ & $21.95_{-4.13}^{+2.96}$ & $3.06_{-0.01}^{+0.01}$ & $1.51_{-0.02}^{+0.02}$ & 1104/895 \\
P050523701402 & 60027.6 & $2.51_{-0.03}^{+0.05}$ & $0.65_{-0.03}^{+0.07}$ & $1.44_{-0.03}^{+0.02}$ & $26.89_{-5.52}^{+7.50}$ & $2.66_{-0.02}^{+0.02}$ & $1.00_{-0.03}^{+0.03}$ & 904/895 \\
P050523701701 & 60031.7 & $2.70_{-0.01}^{+0.01}$ & $1.00_{-0.01}^{+0.0}$ & $1.57_{-0.01}^{+0.01}$ & $23.50_{-4.21}^{+3.61}$ & $2.91_{-0.01}^{+0.01}$ & $1.30_{-0.01}^{+0.01}$ & 1026/895 \\
P050523701703 & 60032.0 & $2.69_{-0.01}^{+0.01}$ & $1.00_{-0.02}^{+0.0}$ & $1.56_{-0.01}^{+0.01}$ & $24.12_{-4.35}^{+4.10}$ & $2.97_{-0.01}^{+0.01}$ & $1.35_{-0.02}^{+0.02}$ & 1021/895 \\
P050523701802 & 60034.4 & $2.57_{-0.01}^{+0.04}$ & $1.00_{-0.02}^{+0.0}$ & $1.52_{-0.01}^{+0.01}$ & $25.31_{-4.04}^{+4.18}$ & $2.92_{-0.01}^{+0.01}$ & $1.54_{-0.02}^{+0.02}$ & 1111/895 \\
P050523701901 & 60035.9 & $2.61_{-0.05}^{+0.07}$ & $0.46_{-0.04}^{+0.06}$ & $1.46_{-0.02}^{+0.01}$ & $27.02_{-5.03}^{+5.79}$ & $2.81_{-0.01}^{+0.01}$ & $0.65_{-0.02}^{+0.02}$ & 893/895 \\
P050523701903 & 60036.2 & $2.51_{-0.08}^{+0.06}$ & $0.37_{-0.04}^{+0.03}$ & $1.48_{-0.02}^{+0.01}$ & $25.89_{-4.94}^{+5.19}$ & $2.75_{-0.01}^{+0.01}$ & $0.58_{-0.02}^{+0.02}$ & 870/895 \\
P050523702001 & 60037.7 & $2.35_{-0.05}^{+0.05}$ & $0.26_{-0.02}^{+0.02}$ & $1.57_{-0.01}^{+0.01}$ & $23.85_{-3.26}^{+3.08}$ & $2.94_{-0.01}^{+0.01}$ & $0.55_{-0.01}^{+0.01}$ & 933/895 \\
P050523702101 & 60039.5 & $2.29_{-0.05}^{+0.06}$ & $0.20_{-0.02}^{+0.02}$ & $1.54_{-0.01}^{+0.01}$ & $23.49_{-3.26}^{+3.70}$ & $2.66_{-0.01}^{+0.01}$ & $0.41_{-0.02}^{+0.02}$ & 882/895 \\
P050523702201 & 60042.8 & $2.29_{-0.05}^{+0.07}$ & $0.18_{-0.01}^{+0.02}$ & $1.50_{-0.01}^{+0.01}$ & $24.13_{-2.92}^{+2.92}$ & $2.51_{-0.01}^{+0.01}$ & $0.36_{-0.02}^{+0.02}$ & 911/895 \\
P050523702301 & 60043.6 & $2.35_{-0.06}^{+0.01}$ & $0.23_{-0.02}^{+0.01}$ & $1.52_{-0.01}^{+0.01}$ & $24.22_{-3.89}^{+2.75}$ & $2.65_{-0.01}^{+0.01}$ & $0.43_{-0.01}^{+0.01}$ & 869/895 \\
P050523702402 & 60046.6 & $2.63_{-0.03}^{+0.02}$ & $0.63_{-0.03}^{+0.02}$ & $1.27_{-0.01}^{+0.01}$ & $31.79_{-5.81}^{+5.69}$ & $2.23_{-0.01}^{+0.01}$ & $0.72_{-0.01}^{+0.02}$ & 1175/895 \\
P050523702501 & 60048.7 & $2.47_{-0.02}^{+0.03}$ & $0.34_{-0.01}^{+0.02}$ & $1.39_{-0.01}^{+0.01}$ & $28.26_{-3.30}^{+4.21}$ & $2.53_{-0.01}^{+0.01}$ & $0.53_{-0.01}^{+0.01}$ & 1225/895 \\
P050523702601 & 60050.5 & $2.20_{-0.06}^{+0.06}$ & $0.19_{-0.02}^{+0.02}$ & $1.45_{-0.01}^{+0.01}$ & $26.04_{-4.59}^{+3.89}$ & $2.53_{-0.01}^{+0.01}$ & $0.45_{-0.02}^{+0.02}$ & 1014/895 \\
P050523702701 & 60052.7 & $2.25_{-0.06}^{+0.07}$ & $0.23_{-0.02}^{+0.02}$ & $1.45_{-0.01}^{+0.01}$ & $25.95_{-4.40}^{+4.32}$ & $2.55_{-0.01}^{+0.01}$ & $0.49_{-0.02}^{+0.02}$ & 1136/895 \\
P050523703301 & 60054.7 & $2.16_{-0.07}^{+0.02}$ & $0.18_{-0.02}^{+0.01}$ & $1.47_{-0.01}^{+0.01}$ & $24.26_{-3.83}^{+2.53}$ & $2.34_{-0.01}^{+0.01}$ & $0.42_{-0.01}^{+0.01}$ & 967/895 \\
P050523703401 & 60055.8 & $2.00_{-0.09}^{+0.08}$ & $0.09_{-0.01}^{+0.01}$ & $1.43_{-0.01}^{+0.01}$ & $23.41_{-3.23}^{+3.19}$ & $1.99_{-0.01}^{+0.01}$ & $0.24_{-0.02}^{+0.02}$ & 938/895 \\
P050523703602 & 60060.5 & $2.05_{-0.10}^{+0.10}$ & $0.13_{-0.02}^{+0.02}$ & $1.47_{-0.01}^{+0.01}$ & $21.71_{-3.83}^{+3.57}$ & $1.87_{-0.01}^{+0.01}$ & $0.28_{-0.02}^{+0.02}$ & 918/895 \\
P050523703901 & 60066.0 & $1.62_{-0.09}^{+0.17}$ & $0.01_{-0.01}^{+0.02}$ & $1.36_{-0.03}^{+0.02}$ & $23.29_{-6.61}^{+7.42}$ & $1.58_{-0.08}^{+0.08}$ & $0.28_{-0.09}^{+0.12}$ & 912/895 \\
    \hline
        \label{HXMTparadisk} &     

    \end{tabular}
}
\end{table*}

\begin{table*}[]
    \centering
		\caption{Fitting results of Insight-HXMT for Model M2. $\Gamma$ is the low-energy power-law photon index, Cov\_frac the coverage factor, $M_{\rm dd}$ the “effective” mass accretion rate of the disk in units of $10^{18}$ g/s,  $f_{\rm col}$ the spectral hardening factor and 
        $T_{\rm eff}$ the effective temperature of the inner zone of the disk, obtained by $T_{\rm in}/f_{\rm col}$, and $R_{\rm in}$ the inner radius of the disk.  }
          \resizebox{\textwidth}{!}{%
		\begin{tabular}{ccccccccc}
		  \hline
		   \hline
          Insight-HXMT & MJD&$\Gamma$& Cov\_frac&$M_{\rm dd}$ &$f_{\rm col}$ &$T_{\rm eff}$&$R_{\rm in}$&$\chi^2$/dof
         \\ObsID &&&&&&(keV)&(km)&
       \\ \hline

P040426302301 & 59803.6 & $1.78_{-0.24}^{+0.27}$ & $0.04_{-0.01}^{+0.02}$ & $1.62_{-0.02}^{+0.02}$ & $1.78_{-0.03}^{+0.79}$ & $0.80_{-0.02}^{+0.02}$ & $26.17_{-0.73}^{+0.79}$ & 996/896 \\  
P040426302403 & 59807.4 & $2.40_{-0.28}^{+0.31}$ & $0.10_{-0.03}^{+0.05}$ & $1.71_{-0.02}^{+0.02}$ & $1.76_{-0.03}^{+1.07}$ & $0.79_{-0.02}^{+0.02}$ & $26.64_{-0.71}^{+1.07}$ & 977/896 \\  
P040426302501 & 59809.5 & $1.84_{-0.17}^{+0.19}$ & $0.06_{-0.01}^{+0.02}$ & $1.79_{-0.01}^{+0.01}$ & $1.8_{-0.02}^{+0.52}$ & $0.8_{-0.01}^{+0.01}$ & $26.795_{-0.543}^{+0.52}$ & 1021/896 \\  
P040426302701 & 59816.2 & $1.63_{-0.21}^{+0.23}$ & $0.03_{-0.01}^{+0.01}$ & $1.94_{-0.01}^{+0.01}$ & $1.84_{-0.02}^{+0.62}$ & $0.81_{-0.01}^{+0.01}$ & $27.12_{-0.32}^{+0.62}$ & 1034/896 \\  
P040426302802 & 59820.5 & $2.19_{-0.30}^{+0.40}$ & $0.06_{-0.02}^{+0.04}$ & $1.94_{-0.02}^{+0.01}$ & $1.82_{-0.02}^{+0.63}$ & $0.80_{-0.01}^{+0.01}$ & $27.71_{-0.57}^{+0.63}$ & 920/896 \\  
P040426302901 & 59822.8 & $1.52_{-0.33}^{+0.41}$ & $0.03_{-0.01}^{+0.02}$ & $1.98_{-0.02}^{+0.02}$ & $1.83_{-0.03}^{+0.61}$ & $0.81_{-0.01}^{+0.02}$ & $27.43_{-1.00}^{+0.61}$ & 869/896 \\  
P040426303001 & 59823.4 & $1.90_{-0.26}^{+0.28}$ & $0.04_{-0.01}^{+0.02}$ & $1.97_{-0.01}^{+0.01}$ & $1.71_{-0.01}^{+0.42}$ & $0.81_{-0.01}^{+0.01}$ & $26.90_{-0.49}^{+0.42}$ & 1020/896 \\  
P051435300101 & 59843.2 & $2.37_{-0.15}^{+0.16}$ & $0.14_{-0.03}^{+0.04}$ & $1.91_{-0.02}^{+0.01}$ & $1.69_{-0.02}^{+0.56}$ & $0.81_{-0.01}^{+0.02}$ & $26.92_{-0.68}^{+0.56}$ & 898/896 \\  
P051435300301 & 59846.2 & $2.57_{-0.14}^{+0.15}$ & $0.17_{-0.03}^{+0.04}$ & $1.93_{-0.01}^{+0.01}$ & $1.73_{-0.02}^{+0.53}$ & $0.80_{-0.01}^{+0.01}$ & $27.47_{-0.48}^{+0.53}$ & 1000/896 \\  
P051435300401 & 59849.1 & $2.75_{-0.15}^{+0.15}$ & $0.27_{-0.05}^{+0.06}$ & $1.74_{-0.02}^{+0.01}$ & $1.77_{-0.02}^{+0.67}$ & $0.79_{-0.01}^{+0.01}$ & $27.52_{-0.47}^{+0.67}$ & 962/896 \\  
P050523700101 & 60003.8 & $2.30_{-0.04}^{+0.06}$ & $0.24_{-0.01}^{+0.02}$ & $3.50_{-0.02}^{+0.01}$ & $1.68_{-0.01}^{+0.34}$ & $0.92_{-0.01}^{+0.01}$ & $28.37_{-0.32}^{+0.34}$ & 931/896 \\  
P050523700201 & 60005.1 & $2.50_{-0.04}^{+0.05}$ & $0.69_{-0.05}^{+0.06}$ & $3.41_{-0.04}^{+0.04}$ & $1.55_{-0.04}^{+1.13}$ & $0.94_{-0.03}^{+0.03}$ & $27.52_{-1.04}^{+1.13}$ & 942/896 \\  
P050523700402 & 60009.3 & $2.70_{-0.04}^{+0.04}$ & $0.78_{-0.06}^{+0.07}$ & $3.13_{-0.05}^{+0.05}$ & $1.49_{-0.03}^{+1.01}$ & $0.91_{-0.02}^{+0.03}$ & $27.95_{-1.07}^{+1.01}$ & 839/896 \\  
P050523700501 & 60011.3 & $2.60_{-0.03}^{+0.04}$ & $0.78_{-0.05}^{+0.07}$ & $3.16_{-0.04}^{+0.03}$ & $1.47_{-0.03}^{+1.15}$ & $0.92_{-0.03}^{+0.02}$ & $27.69_{-0.76}^{+1.15}$ & 827/896 \\  
P050523700901 & 60020.8 & $2.54_{-0.03}^{+0.03}$ & $0.64_{-0.05}^{+0.05}$ & $3.20_{-0.03}^{+0.03}$ & $1.46_{-0.02}^{+0.98}$ & $0.92_{-0.02}^{+0.02}$ & $27.66_{-0.67}^{+0.98}$ & 752/896 \\  
P050523700903 & 60021.1 & $2.57_{-0.05}^{+0.06}$ & $0.55_{-0.04}^{+0.06}$ & $3.36_{-0.03}^{+0.03}$ & $1.57_{-0.02}^{+0.56}$ & $0.91_{-0.02}^{+0.01}$ & $28.65_{-0.58}^{+0.56}$ & 826/896 \\  
P050523701102 & 60022.3 & $2.55_{-0.05}^{+0.04}$ & $0.68_{-0.06}^{+0.07}$ & $3.16_{-0.04}^{+0.04}$ & $1.49_{-0.03}^{+1.05}$ & $0.90_{-0.02}^{+0.03}$ & $29.13_{-1.50}^{+1.05}$ & 928/896 \\  
P050523701302 & 60025.6 & $2.64_{-0.01}^{+0.02}$ & $1.00_{-0.03}^{+0.00}$ & $3.38_{-0.01}^{+0.02}$ & $1.78_{-0.01}^{+0.28}$ & $0.93_{-0.01}^{+0.01}$ & $28.51_{-0.51}^{+0.28}$ & 872/896 \\  
P050523701402 & 60027.6 & $2.45_{-0.05}^{+0.05}$ & $0.56_{-0.05}^{+0.05}$ & $3.10_{-0.03}^{+0.03}$ & $1.58_{-0.03}^{+1.12}$ & $0.91_{-0.03}^{+0.02}$ & $27.52_{-0.63}^{+1.12}$ & 911/896 \\  
P050523701701 & 60031.7 & $2.72_{-0.02}^{+0.01}$ & $1.00_{-0.04}^{+0.00}$ & $3.22_{-0.01}^{+0.02}$ & $1.70_{-0.01}^{+0.35}$ & $0.92_{-0.01}^{+0.01}$ & $27.84_{-0.45}^{+0.35}$ & 857/896 \\  
P050523701703 & 60032.0 & $2.72_{-0.02}^{+0.01}$ & $1.00_{-0.03}^{+0.00}$ & $3.29_{-0.01}^{+0.02}$ & $1.68_{-0.01}^{+0.43}$ & $0.93_{-0.01}^{+0.01}$ & $27.91_{-0.46}^{+0.43}$ & 886/896 \\  
P050523701802 & 60034.4 & $2.59_{-0.0}^{+0.01}$ &
$1.00_{-0.03}^{+0.00}$ & $3.24_{-0.01}^{+0.03}$ & $1.63_{-0.01}^{+0.43}$ & $0.93_{-0.01}^{+0.02}$ & $27.57_{-0.36}^{+0.43}$ & 899/896 \\  
P050523701901 & 60035.9 & $2.39_{-0.05}^{+0.08}$ & $0.30_{-0.03}^{+0.03}$ & $3.35_{-0.02}^{+0.02}$ & $1.60_{-0.02}^{+0.66}$ & $0.91_{-0.02}^{+0.01}$ & $28.36_{-0.51}^{+0.66}$ & 882/896 \\  
P050523701903& 60036.2 & $2.28_{-0.08}^{+0.11}$ & $0.24_{-0.03}^{+0.04}$ & $3.28_{-0.04}^{+0.02}$ & $1.62_{-0.02}^{+0.58}$ & $0.91_{-0.02}^{+0.01}$ & $27.86_{-0.53}^{+0.58}$ & 878/896 \\  
P050523702001 & 60037.7 & $2.06_{-0.05}^{+0.06}$ & $0.15_{-0.01}^{+0.01}$ & $3.49_{-0.01}^{+0.01}$ & $1.70_{-0.01}^{+0.241}$ & $0.92_{-0.01}^{+0.01}$ & $28.26_{-0.27}^{+0.24}$ & 1045/896 \\  
P050523702101& 60039.5 & $2.03_{-0.07}^{+0.07}$ & $0.12_{-0.01}^{+0.01}$ & $3.17_{-0.01}^{+0.01}$ & $1.70_{-0.01}^{+0.35}$ & $0.91_{-0.01}^{+0.01}$ & $27.83_{-0.27}^{+0.35}$ & 1087/896 \\  
P050523702201 & 60042.8 & $2.05_{-0.06}^{+0.08}$ & $0.12_{-0.01}^{+0.02}$ & $3.01_{-0.01}^{+0.01}$ & $1.66_{-0.01}^{+0.21}$ & $0.90_{-0.01}^{+0.01}$ & $27.26_{-0.21}^{+0.21}$ & 1007/896 \\  
P050523702301& 60043.6 & $2.09_{-0.06}^{+0.05}$ & $0.14_{-0.01}^{+0.01}$ & $3.16_{-0.01}^{+0.01}$ & $1.67_{-0.01}^{+0.18}$ & $0.91_{-0.01}^{+0.01}$ & $27.69_{-0.36}^{+0.18}$ & 1007/896 \\  
P050523702402 & 60046.6 & $2.46_{-0.03}^{+0.04}$ & $0.42_{-0.02}^{+0.03}$ & $2.70_{-0.02}^{+0.01}$ & $1.47_{-0.01}^{+0.46}$ & $0.86_{-0.01}^{+0.01}$ & $28.16_{-0.48}^{+0.46}$ & 925/896 \\  
P050523702501& 60048.7 & $2.28_{-0.05}^{+0.04}$ & $0.23_{-0.02}^{+0.02}$ & $3.04_{-0.01}^{+0.01}$ & $1.53_{-0.01}^{+0.31}$ & $0.91_{-0.01}^{+0.01}$ & $27.12_{-0.19}^{+0.31}$ & 1042/896 \\  
P050523702601 & 60050.5 & $1.96_{-0.05}^{+0.09}$ & $0.12_{-0.01}^{+0.02}$ & $3.04_{-0.02}^{+0.01}$ & $1.60_{-0.01}^{+0.31}$ & $0.91_{-0.01}^{+0.01}$ & $27.33_{-0.43}^{+0.31}$ & 960/896 \\  
P050523702701 & 60052.7 & $2.00_{-0.06}^{+0.06}$ & $0.14_{-0.01}^{+0.01}$ & $3.07_{-0.01}^{+0.01}$ & $1.61_{-0.01}^{+0.39}$ & $0.90_{-0.01}^{+0.01}$ & $27.58_{-0.40}^{+0.39}$ & 1003/896 \\  
P050523703301 & 60054.7 & $1.93_{-0.05}^{+0.06}$ & $0.12_{-0.01}^{+0.01}$ & $2.81_{-0.01}^{+0.01}$ & $1.66_{-0.01}^{+0.15}$ & $0.89_{-0.01}^{+0.01}$ & $27.41_{-0.35}^{+0.15}$ & 991/896 \\
P050523703401 & 60055.8 & $1.72_{-0.08}^{+0.10}$ & $0.05_{-0.00}^{+0.01}$ & $2.41_{-0.01}^{+0.01}$ & $1.67_{-0.01}^{+0.25}$ & $0.86_{-0.01}^{+0.01}$ & $26.77_{-0.26}^{+0.25}$ & 916/896 \\  
P050523703602 & 60060.5 & $1.85_{-0.09}^{+0.10}$ & $0.09_{-0.01}^{+0.01}$ & $2.25_{-0.01}^{+0.01}$ & $1.74_{-0.02}^{+0.39}$ & $0.85_{-0.01}^{+0.01}$ & $26.94_{-0.44}^{+0.39}$ & 1052/896 \\  
P050523703901 & 60066.0 & $1.50_{-0.12}^{+0.18}$ & $0.05_{-0.01}^{+0.01}$ & $1.94_{-0.03}^{+0.03}$ & $1.63_{-0.04}^{+1.37}$ & $0.83_{-0.03}^{+0.02}$ & $25.37_{-1.10}^{+1.37}$ & 945/896 \\  
    \hline
        \label{HXMTparakerrbb} &     

    \end{tabular}
}
\end{table*}

\begin{table*}[]
    \centering
		\caption{\textbf{Fitting results of Insight-HXMT for Model M3. $\Gamma$ is the low-energy power-law photon index, $T_{\rm in}$ the temperature of the inner disk,  Cov\_frac the coverage factor,  and $R_{\rm in}$ the inner radius of the disk, $p$ the exponent of the radial dependence of the disk temperature, $F_{\rm disk}$ and $F_{\rm non-thermal}$ are disk flux and non-thermal flux, respectively.  }}
  \resizebox{\textwidth}{!}{%
		\begin{tabular}{ccccccccccc}
		  \hline
		   \hline
          Insight-HXMT & MJD&$\Gamma$& Cov\_frac&$T_{\rm in}$ &$R_{\rm in}$ &$p$&$F_{\rm disk}$&$F_{\rm non-thermal}$&$\chi^2$/dof
         \\ObsID &&&&(keV)&(km)&&($10^{-8}$~erg~s$^{-1}$~cm$^{-2}$)&($10^{-8}$~erg~s$^{-1}$~cm$^{-2}$)&
       \\ \hline

P040426302301 & 59803.6 & $2.40_{-0.38}^{+0.26}$ & $0.11_{-0.04}^{+0.04}$ & $1.30_{-0.02}^{+0.03}$ & $30.79_{-13.47}^{+5.04}$ & $1.00_{-0.06}^{+0.00}$ & $1.21_{-0.01}^{+0.01}$ & $0.09_{-0.02}^{+0.02}$ & 935/894 \\ 
P040426302403 & 59807.4 & $3.04_{-0.16}^{+0.23}$ & $0.28_{-0.05}^{+0.07}$ & $1.29_{-0.01}^{+0.02}$ & $32.17_{-12.89}^{+6.65}$ & $0.99_{-0.08}^{+0.01}$ & $1.29_{-0.02}^{+0.02}$ & $0.11_{-0.02}^{+0.02}$ & 929/894 \\ 
P040426302501& 59809.5 & $2.30_{-0.15}^{+0.17}$ & $0.13_{-0.02}^{+0.03}$ & $1.39_{-0.01}^{+0.01}$ & $23.88_{-7.42}^{+8.29}$ & $0.83_{-0.02}^{+0.02}$ & $1.42_{-0.02}^{+0.02}$ & $0.14_{-0.02}^{+0.03}$ & 925/894 \\ 
P040426302701 & 59816.2 & $2.41_{-0.19}^{+0.15}$ & $0.13_{-0.03}^{+0.03}$ & $1.40_{-0.02}^{+0.02}$ & $27.22_{-8.46}^{+12.12}$ & $0.92_{-0.02}^{+0.06}$ & $1.50_{-0.01}^{+0.01}$ & $0.13_{-0.02}^{+0.02}$ & 872/894 \\ 
P040426302802 & 59820.5 & $2.88_{-0.11}^{+0.28}$ & $0.20_{-0.03}^{+0.07}$ & $1.47_{-0.02}^{+0.01}$ & $19.73_{-5.19}^{+6.77}$ & $0.74_{-0.02}^{+0.02}$ & $1.61_{-0.01}^{+0.01}$ & $0.12_{-0.02}^{+0.02}$ & 872/894 \\ 
P040426302901 & 59822.8 & $2.46_{-0.21}^{+0.21}$ & $0.13_{-0.03}^{+0.04}$ & $1.40_{-0.03}^{+0.02}$ & $25.68_{-7.33}^{+12.97}$ & $0.87_{-0.02}^{+0.08}$ & $1.54_{-0.01}^{+0.01}$ & $0.12_{-0.02}^{+0.03}$ & 824/894 \\ 
P040426303001 & 59823.4 & $2.40_{-0.16}^{+0.20}$ & $0.10_{-0.02}^{+0.03}$ & $1.42_{-0.02}^{+0.04}$ & $20.31_{-9.01}^{+8.25}$ & $0.72_{-0.04}^{+0.03}$ & $1.65_{-0.01}^{+0.01}$ & $0.10_{-0.02}^{+0.02}$ & 965/894 \\ 
P051435300101 & 59843.2 & $2.52_{-0.07}^{+0.14}$ & $0.18_{-0.02}^{+0.04}$ & $1.46_{-0.03}^{+0.02}$ & $17.29_{-5.10}^{+7.21}$ & $0.66_{-0.02}^{+0.03}$ & $1.69_{-0.01}^{+0.01}$ & $0.18_{-0.02}^{+0.02}$ & 895/894 \\ 
P051435300301 & 59846.2 & $2.79_{-0.09}^{+0.07}$ & $0.26_{-0.04}^{+0.03}$ & $1.48_{-0.04}^{+0.02}$ & $17.33_{-4.94}^{+9.07}$ & $0.67_{-0.01}^{+0.03}$ & $1.69_{-0.01}^{+0.01}$ & $0.18_{-0.01}^{+0.01}$ & 999/894 \\ 
P051435300401 & 59849.1 & $2.96_{-0.07}^{+0.15}$ & $0.40_{-0.03}^{+0.08}$ & $1.44_{-0.03}^{+0.02}$ & $18.08_{-5.84}^{+6.82}$ & $0.70_{-0.02}^{+0.02}$ & $1.50_{-0.01}^{+0.01}$ & $0.20_{-0.01}^{+0.01}$ & 962/894 \\ 
P050523700101 & 60003.8 & $2.49_{-0.03}^{+0.04}$ & $0.34_{-0.02}^{+0.02}$ & $1.57_{-0.02}^{+0.01}$ & $23.11_{-5.75}^{+6.34}$ & $0.73_{-0.01}^{+0.01}$ & $3.03_{-0.01}^{+0.01}$ & $0.60_{-0.01}^{+0.01}$ & 808/894 \\ 
P050523700201 & 60005.1 & $2.54_{-0.02}^{+0.04}$ & $0.80_{-0.04}^{+0.05}$ & $1.40_{-0.02}^{+0.02}$ & $33.11_{-9.85}^{+11.91}$ & $0.82_{-0.02}^{+0.03}$ & $2.90_{-0.03}^{+0.03}$ & $1.26_{-0.04}^{+0.04}$ & 924/894 \\ 
P050523700402 & 60009.3 & $2.72_{-0.03}^{+0.03}$ & $0.90_{-0.06}^{+0.05}$ & $1.38_{-0.02}^{+0.02}$ & $28.68_{-8.21}^{+9.72}$ & $0.72_{-0.02}^{+0.03}$ & $2.72_{-0.02}^{+0.02}$ & $1.10_{-0.03}^{+0.03}$ & 833/894 \\ 
P050523700501 & 60011.3 & $2.62_{-0.04}^{+0.03}$ & $0.84_{-0.06}^{+0.06}$ & $1.45_{-0.02}^{+0.02}$ & $23.10_{-7.06}^{+7.76}$ & $0.67_{-0.02}^{+0.02}$ & $2.85_{-0.02}^{+0.02}$ & $1.22_{-0.03}^{+0.03}$ & 848/894 \\ 
P050523700901 & 60020.8 & $2.55_{-0.04}^{+0.03}$ & $0.68_{-0.05}^{+0.04}$ & $1.49_{-0.01}^{+0.03}$ & $20.87_{-6.84}^{+6.28}$ & $0.64_{-0.01}^{+0.01}$ & $2.92_{-0.01}^{+0.01}$ & $1.12_{-0.02}^{+0.02}$ & 764/894 \\ 
P050523700903 & 60021.1 & $2.62_{-0.04}^{+0.05}$ & $0.64_{-0.05}^{+0.06}$ & $1.55_{-0.01}^{+0.02}$ & $20.78_{-6.21}^{+5.63}$ & $0.67_{-0.01}^{+0.01}$ & $3.01_{-0.01}^{+0.01}$ & $0.96_{-0.02}^{+0.02}$ & 840/894 \\ 
P050523701102 & 60022.3 & $2.62_{-0.04}^{+0.04}$ & $0.71_{-0.06}^{+0.06}$ & $1.35_{-0.02}^{+0.01}$ & $31.01_{-9.20}^{+11.22}$ & $0.74_{-0.02}^{+0.04}$ & $2.83_{-0.03}^{+0.03}$ & $1.01_{-0.05}^{+0.05}$ & 935/894 \\ 
P050523701302 & 60025.6 & $2.65_{-0.02}^{+0.01}$ & $1.00_{-0.02}^{+0.00}$ & $2.05_{-0.02}^{+0.02}$ & $10.22_{-2.28}^{+2.28}$ & $0.60_{-0.01}^{+0.01}$ & $3.36_{-0.01}^{+0.02}$ & $1.60_{-0.02}^{+0.03}$ & 878/894 \\ 
P050523701402& 60027.6 & $2.51_{-0.04}^{+0.03}$ & $0.65_{-0.05}^{+0.04}$ & $1.44_{-0.03}^{+0.03}$ & $26.89_{-10.08}^{+12.30}$ & $0.75_{-0.02}^{+0.04}$ & $2.66_{-0.02}^{+0.02}$ & $1.00_{-0.03}^{+0.03}$ & 904/894 \\ 
P050523701701 & 60031.7 & $2.70_{-0.01}^{+0.01}$ & 
$1.00_{-0.02}^{+0.0}$ & $1.78_{-0.02}^{+0.02}$ & $14.46_{-4.01}^{+3.19}$ & $0.64_{-0.01}^{+0.01}$ & $3.09_{-0.01}^{+0.01}$ & $1.37_{-0.01}^{+0.01}$ & 891/894 \\ 
P050523701703 & 60032.0 & $2.71_{-0.02}^{+0.01}$ & $1.00_{-0.02}^{+0.00}$ & $1.87_{-0.01}^{+0.01}$ & $12.30_{-3.04}^{+3.00}$ & $0.61_{-0.01}^{+0.01}$ & $3.23_{-0.01}^{+0.02}$ & $1.44_{-0.02}^{+0.03}$ & 890/894 \\ 
P050523701802 & 60034.4 & $2.61_{-0.03}^{+0.03}$ & $1.00_{-0.04}^{+0.00}$ & $2.12_{-0.03}^{+0.02}$ & $7.99_{-1.92}^{+2.30}$ & $0.56_{-0.03}^{+0.03}$ & $3.35_{-0.01}^{+0.03}$ & $1.73_{-0.03}^{+0.04}$ & 885/894 \\ 
P050523701901 & 60035.9 & $2.50_{-0.05}^{+0.07}$ & $0.38_{-0.03}^{+0.04}$ & $1.55_{-0.02}^{+0.01}$ & $21.03_{-4.79}^{+6.21}$ & $0.68_{-0.01}^{+0.01}$ & $2.96_{-0.01}^{+0.01}$ & $0.66_{-0.02}^{+0.02}$ & 868/894 \\ 
P050523701903 & 60036.2 & $2.42_{-0.05}^{+0.09}$ & $0.31_{-0.02}^{+0.04}$ & $1.56_{-0.02}^{+0.02}$ & $21.10_{-6.28}^{+7.60}$ & $0.69_{-0.01}^{+0.02}$ & $2.87_{-0.01}^{+0.01}$ & $0.59_{-0.02}^{+0.02}$ & 856/894 \\ 
P050523702001 & 60037.7 & $2.28_{-0.04}^{+0.06}$ & $0.23_{-0.01}^{+0.02}$ & $1.62_{-0.02}^{+0.01}$ & $20.79_{-5.30}^{+5.99}$ & $0.71_{-0.01}^{+0.01}$ & $3.03_{-0.01}^{+0.01}$ & $0.56_{-0.02}^{+0.02}$ & 911/894 \\ 
P050523702101 & 60039.5 & $2.23_{-0.05}^{+0.09}$ & $0.18_{-0.01}^{+0.02}$ & $1.58_{-0.01}^{+0.02}$ & $20.80_{-6.89}^{+5.28}$ & $0.71_{-0.02}^{+0.01}$ & $2.73_{-0.01}^{+0.01}$ & $0.42_{-0.02}^{+0.02}$ & 946/894 \\ 
P050523702201 & 60042.8 & $2.21_{-0.05}^{+0.07}$ & $0.16_{-0.01}^{+0.02}$ & $1.55_{-0.01}^{+0.01}$ & $20.54_{-5.71}^{+5.37}$ & $0.70_{-0.01}^{+0.01}$ & $2.61_{-0.01}^{+0.01}$ & $0.37_{-0.02}^{+0.02}$ & 952/894 \\ 
P050523702301 & 60043.6 & $2.26_{-0.04}^{+0.05}$ & $0.20_{-0.01}^{+0.01}$ & $1.58_{-0.01}^{+0.01}$ & $20.12_{-4.38}^{+5.26}$ & $0.69_{-0.01}^{+0.01}$ & $2.76_{-0.01}^{+0.01}$ & $0.45_{-0.01}^{+0.01}$ & 920/894 \\ 
P050523702402 & 60046.6 & $2.40_{-0.04}^{+0.03}$ & $0.38_{-0.03}^{+0.02}$ & $1.54_{-0.02}^{+0.02}$ & $15.38_{-4.46}^{+4.21}$ & $0.58_{-0.01}^{+0.01}$ & $2.63_{-0.01}^{+0.01}$ & $0.72_{-0.01}^{+0.01}$ & 909/894 \\ 
P050523702501 & 60048.7 & $2.27_{-0.04}^{+0.04}$ & $0.23_{-0.02}^{+0.02}$ & $1.56_{-0.01}^{+0.01}$ & $17.27_{-3.77}^{+4.04}$ & $0.62_{-0.01}^{+0.01}$ & $2.85_{-0.01}^{+0.01}$ & $0.56_{-0.02}^{+0.02}$ & 1099/894 \\ 
P050523702601 & 60050.5 & $2.01_{-0.07}^{+0.06}$ & $0.13_{-0.02}^{+0.01}$ & $1.58_{-0.02}^{+0.02}$ & $17.76_{-4.69}^{+5.39}$ & $0.64_{-0.01}^{+0.01}$ & $2.78_{-0.01}^{+0.01}$ & $0.49_{-0.02}^{+0.02}$ & 964/894 \\ 
P050523702701 & 60052.7 & $2.02_{-0.07}^{+0.08}$ & $0.15_{-0.01}^{+0.02}$ & $1.61_{-0.02}^{+0.01}$ & $16.84_{-3.67}^{+4.43}$ & $0.63_{-0.01}^{+0.01}$ & $2.85_{-0.01}^{+0.01}$ & $0.54_{-0.02}^{+0.02}$ & 1027/894 \\ 
P050523703301 & 60054.7 & $2.02_{-0.04}^{+0.06}$ & $0.14_{-0.01}^{+0.01}$ & $1.57_{-0.01}^{+0.01}$ & $18.28_{-4.24}^{+4.46}$ & $0.66_{-0.01}^{+0.01}$ & $2.51_{-0.01}^{+0.01}$ & $0.45_{-0.02}^{+0.02}$ & 963/894 \\ 
P050523703401 & 60055.8 & $1.81_{-0.09}^{+0.04}$ & $0.06_{-0.01}^{+0.01}$ & $1.54_{-0.01}^{+0.01}$ & $17.13_{-4.13}^{+4.40}$ & $0.65_{-0.01}^{+0.01}$ & $2.16_{-0.01}^{+0.01}$ & $0.27_{-0.02}^{+0.02}$ & 905/894 \\ 
P050523703602 & 60060.5 & $2.13_{-0.02}^{+0.12}$ & $0.15_{-0.01}^{+0.03}$ & $1.42_{-0.01}^{+0.01}$ & $25.68_{-8.55}^{+7.45}$ & $0.83_{-0.03}^{+0.02}$ & $1.80_{-0.01}^{+0.01}$ & $0.28_{-0.02}^{+0.02}$ & 910/894 \\ 
P050523703901 & 60066.0 & $1.71_{-0.16}^{+0.01}$ & $0.08_{-0.02}^{+0.01}$ & $1.37_{-0.01}^{+0.05}$ & $22.96_{-8.42}^{+7.53}$ & $0.75_{-0.03}^{+0.02}$ & $1.45_{-0.07}^{+0.07}$ & $0.25_{-0.08}^{+0.08}$ & 895/894 \\ 
    \hline
        \label{HXMTparadiskpbb} &     

    \end{tabular}
}

\end{table*}

%





\bibliography{sample631}{}
\bibliographystyle{aasjournal}



\end{document}